\pdfoutput=1

\documentclass[aps,prd,%
10pt,%
onecolumn,%
tightenlines,%
eqsecnum,%
showpacs,%
floats,%
nofootinbib,%
amsfonts,amssymb,amsmath%
]{revtex4-1}

\usepackage{ifpdf}
\ifpdf
\usepackage[pdftex]{graphicx}
\usepackage{epstopdf}
\else
\usepackage{graphicx}
\fi

\graphicspath{{Figures/}}

\usepackage{hyperref}

\newcommand{\ket}[1]{\mbox{$| {#1} \rangle$}}
\newcommand{\bracket}[2]{\mbox{$\langle {#1} \!\mid\! {#2} \rangle$}} 
\newcommand{\ketbra}[2]{\mbox{$| {#1} \rangle\langle {#2} |$}} 
\newcommand{\melt}[3]{\mbox{$\langle {#1} | {#2} | {#3} \rangle$}}

\newcommand{\expct}[1]{\mbox{$\langle {#1} \rangle$}}

\newcommand{\Projsupb}[2]{\mbox{$P^{{#1}}_{\mbox{\scriptsize{${#2}$}}}$}}

\newcommand{\tri}[2]{\mbox{${\rm tr}[ {#1} \rho {#2} ]$}}

\def\IR{\relax{\rm I\kern-.18em R}}

\def\sys{{\mathcal S}}
\def\hilbert{{\mathcal H}}
\def\Hphys{{\mathcal H}_{\mathrm{phys}}}
\def\cell{{\mathcal V}}
\def\LG{{\mathcal L}_G}

\def\es{{}^3{\mathrm e}}
\def\est{{\mathrm e}}

\def\oV{\mathring{V}}

\def\oq{\mathring{q}}

\def\be{\begin{equation}}
\def\ee{\end{equation}}

\def\p{\partial}
\def\tchi{\Psi'}
\def\tpsi{\Psi}
\def\lp{{l}_{p}}
\def\cons{\sqrt{12 \pi G}}
\def\k{\kappa}

\DeclareMathOperator{\erf}{erf}

\begin{document}

\preprint{Preprint \#: PI-QG-191}

\title{Consistent probabilities in Wheeler-DeWitt quantum cosmology}

\author{David A.\ Craig}
\email[]{E-mail: craigda@lemoyne.edu}
\affiliation{%
Department of Physics, Le Moyne College\\
Syracuse, New York, 13214, USA}

\author{Parampreet Singh}
\email[]{E-mail: psingh@phys.lsu.edu}
\affiliation{%
Perimeter Institute for Theoretical Physics\\
Waterloo, Ontario, N2L 2Y5, Canada\\and\\
Department of Physics, Louisiana State University\\
Baton Rouge, Louisiana, 70803, USA}


\begin{abstract}
We give an explicit, rigorous framework for calculating quantum probabilities
in a model theory of quantum gravity.  Specifically, we construct the
decoherence functional for the Wheeler-DeWitt quantization of a flat
Friedmann-Robertson-Walker cosmology with a free, massless, minimally coupled
scalar field, thus providing a complete decoherent histories formulation for
this quantum cosmological model.  The decoherence functional is applied to
study predictions concerning the model's Dirac (relational) observables; the
behavior of semiclassical states and superpositions of such states; and to
study the singular behavior of quantum Wheeler-DeWitt universes.  Within this
framework, rigorous formul\ae\ are given for calculating the corresponding
probabilities from the wave function when those probabilities may be
consistently defined, thus replacing earlier heuristics for interpreting the
wave function of the universe with explicit constructions.  It is shown
according to a rigorously formulated standard, and in a quantum-mechanically
consistent way, that in this quantization these models are generically
singular.  Independent of the choice of state we show that the probability for
these Wheeler-DeWitt quantum universes to ever encounter a singularity is
unity.  In addition, the relation between histories formulations of quantum
theory and relational Dirac observables is clarified.
\end{abstract}

%

\maketitle

\section{Introduction}
\label{sec:intro}

To extract consistent physical predictions from a quantum theory of gravity
describing the whole universe, it is not enough to have a physical Hilbert
space, inner product and self-adjoint physical operators.  One must also have
a coherent framework within which to assign probabilities for the occurrence
of events and phenomena.  When applied to sub-systems of the universe this
problem has a straightforward conventional solution provided by classical
observers external to the sub-system who assign and measure these
probabilities.  The universe as a whole, on the other hand, is a closed
quantum system for which there are no external classical observers performing
measurements \cite{GMH90a,GMH90b,lesH}.  How, then, are quantum probabilities
to be extracted from quantum amplitudes?

An example of the kind of framework required is a generalization of the
consistent or decoherent histories formulation of quantum mechanics developed
by Griffiths \cite{griffiths84,griffiths08}, Omnes
\cite{omnes88a,*omnes88b,*omnes88c,*omnes89,*omnes90,*omnes92,omnes94},
Gell-Mann and Hartle \cite{GMH90a,GMH90b}, and others, and adumbrated,
particularly in the context of quantum gravity and quantum cosmology, by
Hartle \cite{hartle91a,lesH}, Halliwell
\cite{halliwell99,hallithor01,hallithor02,halliwall06,halliwell09}, and others
\cite{hartlemarolf97,CH04,as05}.  The broad outlines of this formalism will be
described in section \ref{sec:gqt} below, and implemented explicitly and in
detail for a simple quantum cosmological model -- which may then serve as a
guide or template for more complex
models -- in subsequent sections.%
\footnote{Some of these results have been reported previously in 
\cite{CS10a,CS10b}.
} %

The necessity for a predictive framework for closed quantum systems
supplementing the usual technical apparatus of quantum theory is particularly
evident in situations in which one desires to extract predictions concerning
correlations between variables, most particularly correlations extended in
time.  Examples of interest include predictions concerning approximately
semiclassical behavior, or other correlations between the values of variables
at different times.

To be specific, it is well understood that it is not always possible to assign
amplitudes to histories of such correlations in a self-consistent manner
unless the interference between the alternatives vanishes, as illustrated by
the classic example of two-slit quantum interference.  Indeed, it is one of
the signature characteristics of quantum theory that not every history that
can be described can be assigned a meaningful probability.
Most, in fact, cannot.  For example, no meaningful probabilities can be
assigned to the histories which specify which slit the particle passed through
before arriving at a position $y$ on the screen \emph{unless} a mechanism
exists to destroy interference between the possible alternative histories
\{(upper slit,$y$),(lower slit,$y$)\} such as a measuring device to determine
which slit the particle passed through, or coupling to an environment, which
carries away -- effectively, creates a record of 
\cite{GMH90a,GMH90b,hartle91a,GMH93,halliwell99} -- 
essentially the same information.%
\footnote{Indeed, in essence the principal role of measurements by external
observers in the traditional Copenhagan approach to prediction in quantum
theory is to supply physical mechanisms which destroy interference among
histories of the measured quantities in the measured subsystem.
} %
Only when there is no interference between independent alternative
outcomes may probabilities be assigned in a consistent manner%
\footnote{In the sense that the appropriate probability sum rules are
satisfied.
} %
to each of those alternatives.

What is needed, then, is an objective measure both of the interference between
alternative histories of a system, and, when that interference vanishes, the
probability of each such alternative history.


Such a measure is provided by the \emph{decoherence functional}, which
measures both the interference between histories in a complete set of
alternative possibilities, and, when meaningful (because interference between
the alternatives vanishes), the probabilities of those histories
\cite{GMH90a,GMH90b,hartle91a,lesH}.  Specification of the state space,
observables, dynamics, and decoherence functional of a system therefore
permits internally consistent physical predictions to be made concerning that
system independent of any notion of external observers, measuring apparatus,
environments, or other ancillary notions that will not be available to a
theory which seeks to describe closed systems such as the universe as a whole.

More generally, because the entire universe is within the domain of the
quantum theory of gravity, \emph{whatever} the final form of this theory it
must of necessity have associated with it either a decoherence functional, or
some other predictive framework which explains how consistent quantum
predictions can be extracted from it.

Moreover, completely apart from the fundamental quantum question of when
quantum probabilities are consistently defined, it has long been understood
that the interpretation of the ``wave of function of the universe'' is itself
perhaps not so clear.  Indeed, there is a long history of debate in the
context of quantum gravity/quantum cosmology of the question, \emph{given} a
``wave function of the universe'', just what should one \emph{do} with it?
(See \cite{halliwell91:qcbu} for one summary and discussions of some specific
proposals.)  In other words, just how does one extract meaningful physical
predictions from a wave function defined on the superspace of gravitational
degrees of freedom, especially given the diffeomorphism invariance of the
underlying theory?  To date, most of the answers to this question have relied
upon heuristic arguments, rather than complete, internally consistent
frameworks as we seek -- on the foundation laid by Hartle -- to supply here.

It is the aim of the present work to supply an example of a complete, explicit
realization of a ``generalized quantum theory''%
\footnote{As Hartle terms it.
} %
that can address both of these questions in quantum gravity, by constructing a
decoherence functional for a simple quantum cosmological model following the
lead of \cite{CH04}.  We thereby supply an example of a quantum gravitational
model in which predictions extracted from the wave function of the universe
are on a firm footing.  The model is a Wheeler-DeWitt quantization of a flat,
Friedmann-Robertson-Walker universe with a massless, minimally coupled scalar
field; a companion work will construct the complete generalized quantum theory
for a loop quantization of the same model \cite{CS10d}.  Together, these
examples provide a template for construction of generalized quantum theories
for more complex quantum gravitational models, serving as a guide to
strategies for tackling some of the technical and conceptual questions and
problems that will arise in any complete predictive quantum theory of gravity.

The Wheeler-DeWitt quantization of the model we consider is developed in
detail in Ref.\ \cite{aps:improved}, where an inner product and physical
Hilbert space were obtained and expectation values of Dirac observables
studied.  In the classical theory, the model has two disjoint solutions for a
given value of the scalar field momentum.  One of these is expanding which
leads to a big bang in the past, and the other is contracting with a big
crunch under future evolution.  In the quantum theory, numerical simulations
using states which are semi-classical at late times show that such states
remain peaked on classical trajectories and the singularity present in the
classical theory is thus not avoided in the quantum theory.  With an
appropriate choice of lapse, the model can also be solved exactly.  One finds
that arbitrary states which can be interpreted as describing an expanding
universe have zero expectation value of the volume observable in the past
\cite{acs:slqc}.  Similarly, arbitrary states describing contracting universes
lead to zero volume in the future.
Though these studies reveal much about the quantization of the model and the
resulting physical behavior, including the persistence of the classical
singularity for generic expanding or contracting quantum states, important
questions concerning quantum \emph{histories} and their probabilities have so
far remained unaddressed.  Here we answer some of these questions.  In
particular, we construct a decoherence functional for the model universe and
demonstrate how the prior results may be understood and refined within a
consistent framework for quantum prediction.  By way of example we address the
question of whether this Wheeler-DeWitt quantized universe is ever singular.
We show that the probability the quantum universe is singular is unity for
arbitrary states in the physical Hilbert space.  Of some special interest is
the case where the quantum state is a \emph{superposition} of contracting and
expanding universes.  Such a state might be regarded as an analog of a
``Schr\"{o}dinger's cat'' state in ordinary quantum mechanics -- a quantum
superposition of macroscopically distinct states.  A careful consistent
histories analysis reveals that even for such states the probability the
universe avoids the singularity is zero, and a naive expectation that such a
state may avoid the classical singularity turns out to be incorrect.

The outline of the paper is as follows.  Section \ref{sec:gqt} describes the
overall framework of generalized quantum theory.  In section
\ref{sec:WdWmodel} we give details of the cosmological model and its
(canonical) quantization, including a discussion of the model's observables.
Section \ref{sec:dWdW} details the construction of the decoherence functional,
including construction of class operators (histories) and the corresponding
branch wave functions, and discusses the relationship between histories-based
theories and relational observables.  Section \ref{sec:app} runs through a
selection of physical applications:
volume (section \ref{sec:volume}); %
momentum of the scalar field (section \ref{sec:pphi}); %
the behavior of semiclassical states (section \ref{sec:semiclassical}); %
and finally, we give in section \ref{sec:singular} a detailed discussion of
the behavior of these universes near the classical singularity.  In
particular, we show rigorously that in this quantization the universe is
singular for all states -- quantum mechanics does not resolve the classical 
singularity.%
\footnote{This result stands in stark contrast to the case of the loop 
quantization of the same model to be described in \cite{CS10d}, in which the 
classical singularity is \emph{resolved} for generic states.
} %
Section \ref{sec:discuss} closes with some discussion of conceptual and 
practical issues.

\section{Generalized ``Consistent Histories'' Quantum Theory}
\label{sec:gqt}

As originally conceived by Hartle \cite{hartle91a}, ``generalized
quantum theory'' is a distillation of the main ideas of the consistent
histories program for quantum mechanics \cite{GMH90a,GMH90b} to its
essential conceptual elements.  The principal virtue of this
abstraction is that it isolates those elements in a manner suitable
for generalization to theories beyond the original domain of the
program, non-relativistic particle mechanics.  Thus the same
fundamental structure may be applied equally well to particle
mechanics as to relativistic field theories or a quantum theory of
gravity, and implemented as naturally in a functional integral
framework as in an operator quantization.

The essential elements of any generalized quantum theory are:

\begin{enumerate}

\item {\em Fine-Grained Histories:} The most refined descriptions of the
universe it is possible to give.  This might be the set of paths in a
functional integral formulation, or the set of time-ordered products of
one-dimensional Heisenberg projections onto physical observables in a Hilbert 
space quantization.

\item {\em Coarse-Grained Histories:} Specification of the physically
meaningful partitions of the set of fine-grained histories.  In a
covariant quantization of gravitation, for example, only
diffeomorphism invariant classes would be expected to be intelligible.
Coarse-grained sets of histories are called ``coarse-grained
histories''.  Such histories correspond to the physically meaningful
questions that may be asked of a system -- ``Which slit did the
particle pass through?''  Note that \emph{most} -- if not all --
physical questions are of this highly coarse-grained character, not
inquiring, in this example, into other details of the particle's
position.  As such, it is the coarse-grained histories for which
quantum mechanics must be able to determine whether probabilities are
meaningful, and if so, what those probabilities are.

\item {\em Decoherence Functional:} The decoherence functional both measures
the quantum mechanical interference between members of a set of alternative
coarse-grained histories, and, when that interference vanishes, determines the
probabilities of each member of that set.  The decoherence functional is a
natural generalization to closed quantum systems of the algebraic notion of
quantum state \cite{IL94,dac97}, and incorporates the system's boundary
conditions.  Sets of histories with negligible interference between all pairs
of members, as measured by the decoherence functional, are said to decohere,
or to be consistent.%
\footnote{We use these terms interchangeably, being aware that some
authors make more refined distinctions that will not be important
here.  Additionally, the notion of decoherence being applied here is
closely related to, but ultimately distinct from, the idea of
environmental decoherence \cite{giulini,schlosshauer07}.  Indeed, from
our point of view environmental decoherence may be regarded as a
particular physical mechanism which leads to the decoherence (in the
present sense) of the corresponding coarse-grained histories.
} %
It is logically consistent to assign probabilities in an exhaustive set of
alternative histories when, and only when, that set is decoherent according to
the system's decoherence functional.  It is the criterion of consistency,
rather than, for example, any notion of ``measurement'', which determines 
the physically meaningful predictions of the theory.

\end{enumerate}

\subsection*{Quantum theory in Hilbert space}
\label{sec:nrqmop}

For purposes of motivation and later comparison, we describe the formulation
of ordinary Hilbert space quantum mechanics in the language of generalized 
quantum theory.

\subsubsection{Histories, class operators, and branch wave functions}
\label{sec:classopsqm}

Consider a quantum theory for a system $\sys$ on a Hilbert space $\hilbert$
with Hamiltonian $H$ and a family of observables $A^{\alpha}$, labelled by
the index $\alpha$, with eigenvalues $a^{\alpha}_k$ (assumed discrete for
notational convenience only.)  Ranges of eigenvalues will be denoted $\Delta
a^{\alpha}_k$.  Projections onto the corresponding eigensubspaces will be
written 
$\Projsupb{\alpha}{a_k}$ and 
$\Projsupb{\alpha}{\Delta a_k}$, where we have suppressed the 
superscript labelling the eigenvalues in order to minimize the notational clutter.

For a given choice of observable $A^{\alpha_i}$ at each time $t_i$, an
exclusive, exhaustive set of fine-grained histories for $\sys$ may be regarded
as the set of sequences of eigenvalues
$\{h\}=\{(a^{\alpha_1}_{k_1},a^{\alpha_2}_{k_2},\dots,a^{\alpha_n}_{k_n})\}$,
where each $k_i$ runs over the full range of the eigenvalues
$a^{\alpha_i}_{k_i}$, corresponding to the family of histories in which
observable $A^{\alpha_i}$ has value $a^{\alpha_i}_{k_i}$ at time $t_i$.  A
different choice of observables $(\alpha_1,\alpha_2,\dots,\alpha_n)$ leads to
a different exclusive, exhaustive family of histories $\{ h \}$.  From the
propagator
\begin{eqnarray}
U(t_i,t_j) & = & e^{-iH(t_i-t_j)/\hbar} \nonumber\\
 & \equiv & U(t_i-t_j) \nonumber\\
\end{eqnarray}
are constructed the Heisenberg projections
\begin{equation}
\Projsupb{\alpha}{a_k}(t) = 
  U^{\dagger}(t) \Projsupb{\alpha}{a_k} U(t).
\label{eq:Hprojqm}
\end{equation}
The fine-grained history $h$ may then be conveniently represented by the 
operator (often called the ``class operator''%
\footnote{Note that the time ordering employed here is opposite to the
original definition of \cite{GMH90a,GMH90b} and that often found
elsewhere in the literature.  This choice, however, is more convenient
in many formul\ae, particularly in the definition of the decoherence
functional.
} %
for the history $h$)
\begin{subequations}
\begin{eqnarray}
C_h &=& \Projsupb{\alpha_1}{a_{k_1}}(t_1)
        \Projsupb{\alpha_2}{a_{k_2}}(t_2) \cdots
        \Projsupb{\alpha_n}{a_{k_n}}(t_n) 
\label{eq:classopdefqm-a}  \\
    &=& U(t_0-t_1)\Projsupb{\alpha_1}{a_{k_1}}
        U(t_1-t_2)\Projsupb{\alpha_2}{a_{k_2}} \cdots
        U(t_{n-1}-t_n)\Projsupb{\alpha_n}{a_{k_n}}U(t_n-t_0) 
\label{eq:classopdefqm-b}
\end{eqnarray}
\label{eq:classopdefqm}%
\end{subequations}
in the sense that the ``branch wave function''
\begin{subequations}
\begin{eqnarray}
\ket{\psi_h} &\equiv&  C_h^{\dagger}\ket{\psi}  \label{eq:bwfdefqm-a} \\
    &=& U(t_0-t_n)\Projsupb{\alpha_n}{a_{k_n}}
        U(t_{n}-t_{n-1})\cdots 
        U(t_2-t_1)\Projsupb{\alpha_1}{a_{k_1}} 
        U(t_1-t_0)\ket{\psi}
\label{eq:bwfdefqm-b}
\end{eqnarray}
\label{eq:bwfdefqm}%
\end{subequations}
constructed from the initial state $\ket{\psi}$ is the (un-normalized) quantum
state for a system which has followed this particular history.%
\footnote{One must be slightly careful with this interpretation.  Because of
the leading factor of $U$ in Eq.\ (\ref{eq:bwfdefqm-b}), $\ket{\psi_h}$ as
defined here is actually the \emph{initial} state that \emph{under normal
Schr\"{o}dinger evolution} (no collapses) from $t_0$ to any $t > t_n$ will
\emph{evolve into} the state that has ``followed'' this particular history up
to $t_n$.  See Section \ref{sec:bwfsolns} for an alternative normalization.
\label{foot:Hbwf}
} %
The projections implement, in the standard Copenhagen/von Neumann way of
thinking, ``wave function collapse''.  From the present point of view,
however, it is more natural to regard $\ket{\psi}$ as ``the state'' of the
system, and the branch wave function merely as a tool from which one may
ultimately construct the probabilities of individual histories.

Note that since $\sum_{k}\Projsupb{\alpha}{a_k}=1$ for each observable
$\alpha$,
\begin{subequations}
\begin{eqnarray}
\sum_h C_h & = & \sum_{k_1}\sum_{k_2}\cdots \sum_{k_n} \, 
                     \Projsupb{\alpha_1}{a_{k_1}}(t_1)
                     \Projsupb{\alpha_2}{a_{k_2}}(t_2) \cdots
                     \Projsupb{\alpha_n}{a_{k_n}}(t_n)
\label{eq:classopsumqm-a}\\
 & = & 1, 
\label{eq:classopsumqm-b}
\end{eqnarray}
\label{eq:classopsumqm}%
\end{subequations}
corresponding to the fact that the set of fine-grained histories $\{ h \}$
represents a mutually exclusive, collectively exhaustive description of the
possible fine-grained histories of $\sys$.  Accordingly,
\begin{subequations}
\begin{eqnarray}
\sum_h \ket{\psi_h} & = & \sum_h C_h^{\dagger} \ket{\psi} \label{eq:bwfsumqm-a}\\
 & = & \ket{\psi}.    \label{eq:bwfsum-b}
\end{eqnarray}
\label{eq:bwfsum}%
\end{subequations}

Coarse-grained histories correspond to coarse-grainings of the projections 
at some or all of the times $t_i$:
\begin{equation}
C_{\overline{h}} = 
\sum_{a_{k_{1}}\in\Delta a_{\bar{k}_1}}
\sum_{a_{k_{2}}\in\Delta a_{\bar{k}_2}}\cdots 
\sum_{a_{k_{n}}\in\Delta a_{\bar{k}_n}} \, 
  \Projsupb{\alpha_1}{a_{k_{1}}}(t_1)
  \Projsupb{\alpha_2}{a_{k_{2}}}(t_2) \cdots
  \Projsupb{\alpha_n}{a_{k_{n}}}(t_n),
\label{eq:classopcgqm}
\end{equation}
where the bar denotes a coarse-graining including the more finely-grained
history $h$.  (It is not intended as a general signifier that $h$ is
coarse-grained.  There is no special notation for that, since most histories
of physical interest will be coarse-grained.)  In general, these operators
will not be simple products of projections as in (\ref{eq:classopdefqm}),
though they can be in simple cases; coarse-grained histories for which they
are, are called ``homogeneous''.  Branch wave functions for coarse-grained
histories are defined as in Eq.\ (\ref{eq:bwfdefqm}), and are obviously simply
sums (superpositions) of the more finely-grained branch wave functions.

\subsubsection{The decoherence functional}
\label{sec:dfqm}

For a pure initial state $\ket{\psi}$, the decoherence functional of standard 
quantum theory is defined as%
\footnote{For mixed initial states $\rho$, the definition is 
$d(h,h')=\tri{h^{\dagger}}{h'}$.  In this form it is especially clear that the 
decoherence functional is a natural generalization of the notion of ``quantum 
state'' as it is employed in algebraic formulations of quantum mechanics 
\cite{dac97}.  
Further generalizations are of course possible, resulting in a ``generalized 
quantum theory'' \cite{hartle91a,lesH}, but the choice given here 
corresponds to quantum mechanics as it is usually done.
} %
\begin{equation}
d(h,h') = \bracket{\psi_{h'}}{\psi_{h}}.
\label{eq:dfdefqm}
\end{equation}
Note that from Eq.\ (\ref{eq:bwfsum}), the decoherence functional is 
normalized:
\begin{subequations}
\begin{eqnarray}
\sum_{h,h'} d(h,h') & = & \bracket{\psi}{\psi} \label{eq:dfnorm-a}\\
                    & = & 1. \label{eq:dfnorm-b}
\end{eqnarray}
\label{eq:dfnorm}%
\end{subequations}

When interference between all the members of an exclusive, exhaustive set of 
coarse-grained histories $\{ h \}$ vanishes,
\begin{equation}
d(h,h') = 0, \qquad h\neq h',
\label{eq:mediumdecoh}
\end{equation}
that set of histories is said to decohere, or be consistent.%
\footnote{Other criteria for decoherence are possible \cite{GMH93}.  The 
condition (\ref{eq:mediumdecoh}), sometimes called ``medium decoherence'', is 
the simplest, and, it appears, most broadly applicable \cite{Diosi04}.
} %
In such sets, the probabilities of the individual histories are then simply 
the diagonal elements of the decoherence functional,
\begin{equation}
p(h) = d(h,h).
\label{eq:dfprobqm}
\end{equation}
It is easily verified that this is simply the standard L\"{u}ders-von Neumann 
formula for probabilities of sequences of outcomes in ordinary quantum 
theory, when such probabilities may be defined -- typically in measurement 
situations.
In the framework of generalized (decoherent histories) quantum theory, 
however, no external notion of observers or measurement is required.  It is 
the criterion (\ref{eq:mediumdecoh}) that determines when probabilities may 
be defined, and which ensures they are meaningful in the sense that 
probability sum rules are obeyed when histories are coarse-grained:
\begin{equation}
p(h_1+h_2) = p(h_1) + p(h_2)
\label{eq:probsumqm}
\end{equation}
in decoherent sets.  Consistent sets are thus characterized by a diagonal
decoherence functional,
\begin{equation}
d(h,h') = p(h)\, \delta_{h'h}.
\label{eq:ffqm}
\end{equation}
From Eqs.\ (\ref{eq:dfnorm}) and (\ref{eq:mediumdecoh}) it is clear
that $\sum_h p(h) = p(1)=1$ in decoherent sets of histories, as
clearly must be the case.

In quantizing simple cosmological models it will turn out to be possible,
following the lead of \cite{hartle91a, lesH,CH04}, to pattern the construction
of their generalized quantum theories in close analogy with that of ordinary
particle mechanics, and many of the above formul\ae\ -- suitably
re-interpreted -- will be taken over wholesale.

\subsubsection{Comment: histories and branch wave functions}
\label{sec:bwfsolns}

A few comments concerning the definitions of class operators and branch wave 
functions are in order.

First, fine-grained histories are sometimes defined without the trailing
factor of $U(t_n-t_0)$ in Eq.\ (\ref{eq:classopdefqm-b}), leading to the
possibly more appealing expression for the branch wave function
\begin{equation}
\mbox{\scriptsize $
\begin{pmatrix}
\text{alternate}\\
\text{definition}  
\end{pmatrix}
$}  \qquad
\ket{\psi_h} = 
\Projsupb{\alpha_n}{a_{k_n}}
U(t_{n}-t_{n-1})\cdots 
U(t_2-t_1)\Projsupb{\alpha_1}{a_{k_1}} 
U(t_1-t_0)\ket{\psi}.
\label{eq:bwfdefqmalt}
\end{equation}
In this form, the branch wave function is precisely the quantum state of a 
system at the moment $t=t_n$ that has followed the history $h$ up to that 
time.  (See footnote \ref{foot:Hbwf}.)
On the other hand, with this choice, fine-grained histories are no longer 
simply products of Heisenberg projections, and no longer sum to unity.  
Rather,
\begin{equation}
\mbox{\scriptsize $
\begin{pmatrix}
\text{alternate}\\
\text{definition}  
\end{pmatrix}
$}  \qquad
\sum_h C_h = U(t_n-t_0).
\label{eq:classopsumqmalt}
\end{equation}
Indeed, this normalization for class operators is rather more natural in
functional integral formulations of quantum theory.  Both conventions appear
in the literature.  Both definitions for fine-grained class operators, though, 
lead to the same decoherence functional, Eq.\ (\ref{eq:dfdefqm}), since they 
differ only by an overall unitary factor.

Second, it should be clearly understood that, as the definitions of branch
wave functions and class operators so far stand, \emph{branch wave functions
are not functions of $t$, and are not solutions of the Schr\"{o}dinger
equation.} Indeed, one would not expect that they could be, since in the usual
picture they represent the state that results from the initial state
$\ket{\psi}$ under unitary evolution \emph{punctuated by discontinuous
``collapses'' onto the} $\Projsupb{\alpha_i}{a_{k_i}}$.

In order to construct from the branch wave functions defined in Eq.\ 
(\ref{eq:bwfdefqm}) a truly \emph{spacetime} wave function, we can define
(returning to our original definitions, Eqs.\ 
(\ref{eq:classopdefqm})-(\ref{eq:bwfdefqm}))
\begin{equation}
C_h(t) = C_h\, U^{\dagger}(t-t_0),
\label{eq:classopdefqmt}
\end{equation}
so that
\begin{subequations}
\begin{eqnarray}
\ket{\psi_h(t)} & = & C_h^{\dagger}(t)\ket{\psi} \label{eq:bwfdefqmt-a}\\
    & = & U(t-t_0) \ket{\psi_h} \label{eq:bwfdefqmt-b}\\
    & = & U(t-t_n)\Projsupb{\alpha_n}{a_{k_n}}
          U(t_{n}-t_{n-1})\cdots 
          U(t_2-t_1)\Projsupb{\alpha_1}{a_{k_1}} 
          U(t_1-t_0)\ket{\psi}.
\label{eq:bwfdefqmt-c}
\end{eqnarray}
\label{eq:bwfdefqmt}%
\end{subequations}
$\ket{\psi_h(t)}$, so defined, is a solution of the Schr\"{o}dinger equation
for all $t$, even for $t<t_n$.  And this choice again clearly leaves the 
definition of the decoherence functional unchanged.  With this definition, 
however, 
\begin{equation}
\sum_h C_h(t) = U(t-t_0)
\label{eq:classopsumqmt}
\end{equation}
rather than the identity, so that
\begin{equation}
\sum_{h} \ket{\psi_h(t)} = \ket{\psi(t)}.
\label{}
\end{equation}

Nonetheless, care must be taken.  $\ket{\psi_h(t)}$ for $t<t_n$ is \emph{not}
in general the same as $\ket{\psi}$ evolved according to the Schr\"{o}dinger
equation punctuated by discontinuous ``collapses'' (projections) at the times
$t_i$ until after the last projection under consideration, $t>t_n$.  As an
extreme example, it will \emph{not} normally be the case that
$\ket{\psi_h(t_0)}=\ket{\psi}$.  Rather, $\ket{\psi_h(t_0)} = \ket{\psi_h}$,
while $\ket{\psi_h(t_n)}$ is in fact the state given in Eq.\
(\ref{eq:bwfdefqmalt}) of a system at $t=t_n$ that has ``followed'' the
history $h$ up to that time, and $\ket{\psi_h(t)}$ the Schr\"{o}dinger evolute
of that state thereafter.  (See footnote \ref{foot:Hbwf}.)

Henceforth, $\ket{\psi_h}$ will wherever appropriate refer to
$\ket{\psi_h(t)}$ unless otherwise noted.

\section{Flat scalar FRW and its Wheeler-DeWitt quantization}
\label{sec:WdWmodel}

In order to implement the decoherent histories program in quantum gravity we
must have models over which there is essentially complete theoretical control:
the physical Hilbert space, including especially the inner product; dynamics,
including the propagator; observables and their spectral decompositions; and
boundary conditions.  Few such models are available.  In this work we study
the Wheeler-DeWitt quantization of a homogeneous and isotropic universe with a
massless, minimally coupled scalar field.  The quantization of this model has
been recently developed in the context of loop quantum cosmology
\cite{aps:improved,acs:slqc}.  In the following we revisit this Wheeler-DeWitt
quantization using (except as indicated) the conventions introduced in these
papers, to which we refer the reader for more details and a comparison to loop
quantum cosmology.  It should be emphasized that even though the model we
consider is simple, it is rich enough to demonstrate some non-trivial aspects
of the quantization of a gravitational theory; the consistent histories
formalism; and the manner in which a probabilistic meaning can be given to the
occurrence of a singularity in the Wheeler-DeWitt theory.  It should also
serve as a valuable guide for the consistent histories formulation of more
complex models when the requisite theoretical control is achieved.

\subsection{Classical homogeneous and isotropic models}
\label{sec:classical}

The metric for homogeneous and isotropic Friedmann-Robertson-Walker (FRW)
cosmologies may be decomposed as
\begin{subequations}
\begin{eqnarray}
g_{ab} & = & -n_a n_b + h_{ab} \label{eq:gFRWa}\\
 & = & -N^2 dt_a dt_b + a^2(t)\oq_{ab}, \label{eq:gFRWb}
\end{eqnarray}
\label{eq:gFRW}%
\end{subequations}
where $t$ is a global time, $n_a = -N\, dt_a$ (so $n^a$ is future-directed)
with lapse $N$, and $\oq_{ab}$ is a fixed (${\mathcal L}_t\oq_{ab}=0$)
fiducial metric on the spatial slices $\Sigma$ -- flat, in the case of $k=0$
FRW spacetime which we consider below -- and $a(t)$ is the scale factor.  If
$\es$ is the volume element associated with $\oq_{ab}$, the physical volume
element (\emph{i.e.\ }that associated with $g_{ab}$) is
\begin{equation}
\sqrt{-g}\, \est = \sqrt{-g}\, dt\wedge\es = N a^3\, dt\wedge\es.
\label{eq:volelt}
\end{equation}

The spatial 3-manifold $\Sigma$ may be topologically $\IR^3$, or closed (such
as a torus.)  If $\Sigma$ is topologically open, we introduce a fixed fiducial
cell $\cell$ to define the symplectic framework.  The cell $\cell$ has volume
$\oV$ relative to $\oq_{ab}$,
\begin{equation}
\int_{\cell}\es = \oV,
\label{eq:fidcellvol}
\end{equation}
and all integrations are restricted to $\cell$.  The choice of the topology of
$\Sigma$ plays little role in our analysis and without any loss of generality
we assume it to be 
$\IR^3$  
and take $\cell$ to have unit volume $\oV$.  Note
that the necessity to introduce the fiducial cell $\cell$ does not arise if we
consider a closed ($k = +1$) universe, in which case $\oV$ is simply the
volume of $\Sigma$ with respect to $\oq_{ab}$.

The gravitational part of the action $S_{\mathrm{grav}} = \int\!\LG/16\pi G$
can be computed from the Lagrangian density%
\footnote{This expression for $\LG$, while giving the correct equations of 
motion and Hamiltonian for these models, leaves out boundary terms which 
either integrate to zero or are cancelled by the Gibbons-Hawking term in the 
full gravitational action. See \cite[section E.2.28]{Wald84}
and subsequent discussion.
} %
\begin{equation}
\LG = \sqrt{h}N[K_{ab}K^{ab}-K^2+{}^3\!R], 
\end{equation}
where $K_{ab}=\frac{1}{2}{\mathcal L}_n h_{ab} = \dot{h}_{ab}/2N = 
(\dot{a}/aN)h_{ab}$ and ${}^3\!R = 6k/a^2$.
%
%
%
%
%
%
For the flat ($k=0$) universe, the gravitational action becomes
\begin{equation}
S_{\mathrm{grav}} = \frac{3}{8 \pi G} \, \int d t (- a \dot a^2),
\end{equation}
where we have carried out the integration over the unit volume of the fiducial
cell, and chosen the lapse $N$ to be equal to unity.
The gravitational phase space variables are the scale factor $a$ and its
conjugate 
\begin{equation}
p_a = -\frac{3}{4\pi G}\, a\dot{a},
\end{equation}
for which 
\begin{equation}
\{a, p_a\} = 1 ~.
\end{equation}
The Hamiltonian $H_{\mathrm{grav}}$ can then be obtained by a Legendre
transformation. The total Hamiltonian can  be written as 
\begin{equation}
H = - \frac{2 \pi G}{3} \, \frac{p_a^2}{a} + {H}_{\mathrm{matt}},
\end{equation}
where ${H}_{\mathrm{matt}}$ is the matter Hamiltonian.  We take the matter to
be a massless, minimally coupled scalar field $\phi$ with action
\begin{equation}
S_{\mathrm{matt}} = \int d t \, a^3 \frac{\dot \phi^2}{2}
\end{equation}
and Hamiltonian
\begin{equation}
{H}_{\mathrm{matt}} = \frac{p_\phi^2}{2 a^3},
\end{equation}
where $p_\phi$ denotes the momentum of the scalar field.  The matter phase
space variables $\phi$ and $p_\phi$ satisfy $ \{\phi,p_\phi\} = 1$, and it is
easy to see using Hamilton's equations that $p_\phi = a^3 \, \dot \phi$.

Physical trajectories are solutions to the Hamiltonian constraint
\begin{equation}
- a^2 \, p_a^2 +  \frac{3}{4 \pi G} \, p_\phi^2 \approx 0.
\label{eq:classconstraintaphi}
\end{equation}

Here, however, we will work with a different set of gravitational
phase space variables introduced in loop quantum cosmology \cite{acs:slqc} and
related to the set $(a,p_a)$ by a canonical transformation.
These are the volume of the universe $\varepsilon V = \varepsilon a^3$ and its
canonical conjugate
\begin{equation}
\beta = -\varepsilon \frac{4\pi G}{3}\frac{p_a}{a^2},
\label{eq:betadef}
\end{equation}
which satisfy%
\footnote{In comparison to the analysis in Ref.\ \cite{acs:slqc}, we have put
the Barbero-Immirzi parameter $\gamma$ of loop quantum gravity equal to unity.
In these equations $\varepsilon = \pm 1$ is an orientation factor that
determines the orientation relative to a fiducial triad in terms of which
$\oq_{ab}$ is expressed.  It will play little obvious role in the analysis of
the Wheeler-DeWitt theory, as will be described below, but its presence is 
necessary for the consistency of the quantization.
} %
\begin{equation}
\{\varepsilon V, \beta\} \, = -4 \pi G.
\end{equation}
Solving the classical Hamiltonian constraint $H \approx 0$,
\begin{equation}\label{eq:hamcons}
- \beta^2 V^2 + \frac{4 \pi G}{3} \, p_\phi^2 \approx 0,
\end{equation}
one finds that $\beta^2$ measures the energy density $\rho = p_\phi^2/2 V^2$
of the universe on classical solutions.  Thus, $\varepsilon V$ and $\beta$ are
convenient phase space variables which capture the volume of the universe and
its spacetime curvature.%
\footnote{Due to the constant equation of state of the scalar matter it is
straightforward to see that energy density, and hence the Hubble rate $\beta$,
is sufficient to completely capture the spacetime curvature.
} %
The complete phase space of the model is labelled by 
$(\varepsilon V,\beta,\phi,p_\phi)$.

\begin{figure}[tbh!]
\includegraphics{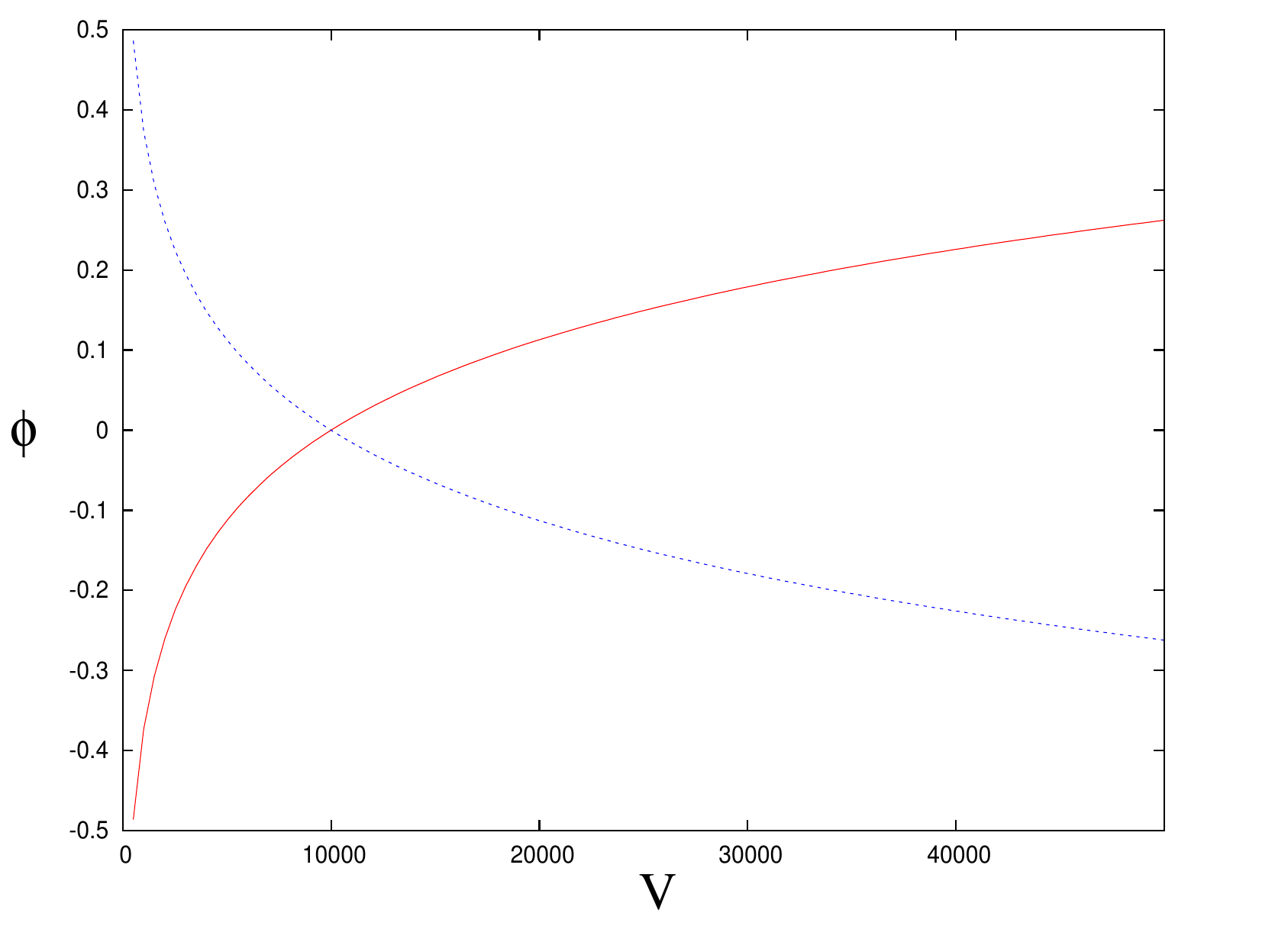}
\caption{Classical trajectories (Eq.\ (\ref{eq:cltraj})) for a massless scalar
field in a flat isotropic universe are shown for positive volume 
($\varepsilon = +1$).  The solid
(red) %
curve corresponds to the expanding branch and the dashed 
(blue) %
curve to the contracting branch.  The branches are disjoint, and singular in
the past and future respectively.}
\label{fig:classicalsolns}
\end{figure}

The dynamical trajectories are obtained by solving Hamilton's equations, 
which show that $p_\phi$ is a constant of the motion, and
\begin{equation}\label{eq:cltraj}
\phi = \pm \, \frac{1}{\cons} \, \ln \left|\frac{V}{V_o}\right| + \phi_o,
\end{equation}
where $V_o$ and $\phi_o$ are constants of integration.  Regarding the scalar
field $\phi$ as an emergent ``clock'', the classical trajectories correspond
to disjoint expanding ($+$) and contracting ($-$) branches.  There exists a
past singularity (the big bang) in the expanding branch as $\phi \rightarrow -
\infty$.  Similarly, the contracting branch faces a big crunch singularity as
$\phi \rightarrow \infty$.  (See Fig.\ \ref{fig:classicalsolns}.)  It is to be
noted that {\it all} classical solutions of this model are singular.

\subsection{Wheeler-DeWitt quantization}
\label{sec:WdWquantization}

The canonical quantization of the classical model proceeds by promoting the
classical phase variables and Poisson brackets to operators and
commutators.  The commutator between gravitational phase variables can be
written as
\begin{equation}
[\varepsilon\hat{V},\hat{\beta}]= -4\pi G\, i \hbar. 
\end{equation}
For convenience, following Ref.\ \cite{aps:improved} we employ the
dimensionless volume $\nu$ given by 
\begin{equation}
\nu :=  \varepsilon \frac{V}{C\, \lp^3}.
\end{equation}
Here $C$ is a dimensionless constant.%
\footnote{In loop quantum cosmology its value is $(4\pi/3)^{(3/2)}/K$, where
$K=2\sqrt{2}/3\sqrt{3\sqrt{3}}$ \cite{aps:improved}, but insofar as this paper
is concerned the value of $C$ is immaterial.   Note we take $\lp = 
\sqrt{G\hbar}$.
} %
In the $\nu,\phi$ representation the action of $\hat \nu$ and $\hat \phi$ is
multiplicative in the kinematical Hilbert space ${\cal H}_{\mathrm{kin}} =
L^2(\IR^2, d \nu d\phi)$, whereas $\hat \beta$ and $\hat p_\phi$ act
as differential operators:
\begin{equation}
\hat \beta = \frac{4\pi}{C\lp} \, i\frac{\p}{\p \nu}        ~~ \mathrm{and} ~~ 
\hat p_\phi = - i \hbar \, \frac{\p}{\p \phi} ~.
\end{equation}

With this action, quantization of the classical constraint Eq.\
(\ref{eq:hamcons}) results in the Wheeler-DeWitt equation (in the volume
representation):
\begin{equation}\label{eq:wdw}
\p_\phi^2 \, \Psi(\nu,\phi) = 12 \pi G \, \nu \p_\nu \, (\nu \p_\nu \, \Psi(\nu,\phi)) 
   =: - \Theta(\nu) \, \Psi(\nu,\phi) ~.
\end{equation}
The choice of factor ordering corresponds to the Laplace-Beltrami operator of
the DeWitt metric on the two dimensional configuration space $(\nu,\phi)$
\cite{aps:improved}.  This will be the most appropriate choice for comparison
with the case of loop quantum cosmology \cite{aps:improved,acs:slqc}.

The Wheeler-DeWitt equation thus takes the form of a Klein-Gordon equation in
a static spacetime.  The scalar field $\phi$ corresponds to the ``time'' and
the operator $\Theta$ corresponds to the spatial Laplacian.  Though the
presence of such a notion of emergent ``time'' helps in the
\emph{interpretation} of various physical results, it should be emphasized
that it is in no way vital to the analysis -- though it will play a role in
the quantum observables we choose to study in the sequel.  Further, given the
simple form of the Wheeler-DeWitt equation for this model one may also regard
$\nu$ as a ``clock'' with respect to which one can measure the evolution of
matter degrees of freedom and the spacetime curvature.


It is important to note that because of the factor of $\varepsilon$ in its
definition, the volume variable $\nu$ ranges over the entire real line,
$-\infty < \nu < \infty$.  However, as discussed in
\cite{aps,aps:improved,acs:slqc}, since the action for the Wheeler-DeWitt
model is invariant under a change of orientation $\varepsilon \rightarrow
-\varepsilon$, physical states $\Psi(\nu,\phi)$ may be taken to be symmetric
in $\nu$.  It suffices, therefore, to restrict attention strictly to 
$\nu \geq 0$, and we will do so everywhere in the sequel.

The operator $\Theta$ is positive definite and self-adjoint on (the symmetric
sector of) $L^2(\IR,\nu^{-1} d \nu)$ \cite{aps:improved}, with eigenfunctions
in the volume representation
\begin{equation}
e_k(\nu) = \frac{1}{\sqrt{2 \pi}} e^{i k  \ln |\nu|} 
\end{equation}
satisfying 
\begin{equation}
\Theta(\nu) \, e_k(\nu) = \omega^2 \, e_k(\nu) ~.
\end{equation}
Here $k$ is a real number for which $\omega = \cons |k|$.
These eigenfunctions form an orthonormal basis,%
\footnote{The normalization chosen here is appropriate to the choice to 
restrict attention to $\nu \geq 0$.
} %
\begin{equation}
\int_{0}^{\infty} \, \frac{d \nu}{\nu} \, \bar e_{k'}(\nu) \, e_k(\nu) = \delta(k - k') ~.
\end{equation}
Any symmetric solution to the Wheeler-DeWitt equation Eq.\ (\ref{eq:wdw}) can
be expanded in terms of the $e_k(\nu)$ and expressed in terms of positive and
negative frequency parts which satisfy the first order version of the quantum
constraint Eq.\ (\ref{eq:wdw}):
\begin{equation}\label{eq:fwdweq}
{\mp} \, i \p_\phi \, \tpsi(\nu,\phi) = \sqrt{\Theta} \, \tpsi(\nu,\phi) ~.
\end{equation}
The positive and negative frequency sectors are disjoint (see below),
resulting in two identical copies of the solution space.  Restricting to the
positive frequency sector, dynamical evolution is given by propagation with
the negative root of Eq.\ (\ref{eq:fwdweq}):
\begin{equation}
U(\phi-\phi_0) = e^{i\sqrt{\Theta}(\phi-\phi_0)} ~.
\label{eq:WdWprop}
\end{equation}

Positive frequency solutions of Eq.\ (\ref{eq:fwdweq}) can be further written
as a combination of ``left-moving'' (contracting) and ``right-moving''
(expanding) components in a plot of $\phi$ \emph{vs.\ }$\nu$.  Evolution with
Eq.\ (\ref{eq:WdWprop}) then leads to
\begin{subequations}
\begin{eqnarray}
\tpsi(\nu,\phi) &=&  
\frac{1}{\sqrt{2 \pi}} \, \int_{-\infty}^{\infty} d k \, \psi(k) \, 
                   e^{ i k  {\ln \nu}} \, e^{i \omega (\phi - \phi_0)} 
\label{eq:quant_traj-a} \\
&=&  
\frac{1}{\sqrt{2 \pi}} \, \int_{-\infty}^0 d k \, \psi(k) \, 
                   e^{i k(\ln \nu - \cons (\phi - \phi_0))} + 
\frac{1}{\sqrt{2 \pi}} \, \int_0^\infty d k \, \psi(k) \, 
                   e^{i k(\ln \nu + \cons (\phi - \phi_0))} \nonumber \\ 
\label{eq:quant_traj-b}\\
&=& \tpsi^R(\nu_-) + \tpsi^L(\nu_+) ~,   
\label{eq:quant_traj-c}
\end{eqnarray}
\label{eq:quant_traj}
\end{subequations}
where $\nu_{\pm}= \ln \nu \pm \cons (\phi - \phi_0)$.  

In order to determine the physical Hilbert space we first consider a set of
Dirac observables -- self-adjoint operators which commute with the constraint
\cite{Rovelli:obs,Marolf:obs,Dittrich:obs}.  One of them is the invariant of
the model, the momentum $p_\phi$ of the scalar field which commutes with the
Hamiltonian.  Another observable which commutes with the Hamiltonian can be
constructed by noting that $\phi$ plays the role of ``time'' in the model in
the sense that it can be used to label the flow of the evolution operator.  If
we then consider a variable such as volume ($\nu$) or energy density ($\rho$),
its value at a fixed value of $\phi = \phi_o$ is an invariant.  Thus, even
though $\nu$ does not commute with the Hamiltonian, the observable $\nu_{|\phi}$ 
is a Dirac observable.%
\footnote{These observables can be defined in a similar way if one chooses
volume to be the ``clock''.  For example one could measure the value of $\phi$
at a given $\nu$: $\phi_{|\nu}$.
} %
In general we can  consider a local operator $\hat{A}$ that does not 
(necessarily) %
commute with the constraint.  The corresponding relational observable 
\begin{subequations}
\begin{eqnarray}
\hat{A}|_{\phi_o}\ket{\Psi(\phi)} & = & 
U(\phi-\phi_o)\hat{A} \ket{\Psi(\phi_o)}
\label{eq:relnlobs-a}\\
& = & 
U(\phi_o-\phi)^{\dagger}\hat{A} U(\phi_o-\phi) \ket{\Psi(\phi)},
\label{eq:relnlobs-b}
\end{eqnarray}
\label{eq:relnlobs}%
\end{subequations}
giving the value of $\hat{A}$ at $\phi=\phi_o$, will then be a Dirac
observable. From Eq.\ (\ref{eq:relnlobs}), note that
\begin{equation}
\hat{A}|_{\phi_o} = U(\phi_o-\phi)^{\dagger}\hat{A} U(\phi_o-\phi),
\label{eq:relnlobsA}
\end{equation}
so that $\hat{A}|_{\phi_o}$ may be thought of as a ``time-reversed''
Heisenberg-picture observable.
Note its action \emph{does} depend on the minisuperspace slice $\phi$ on which
it acts,
so that when we are being careful we should write $\hat{A}|_{\phi^*}(\phi)$.

The action of the Dirac observables preserves the positive and negative
frequency subspaces and as noted it therefore suffices to consider only one of
them to extract physical predictions.  For the space of symmetric positive
frequency solutions this action is given by
\begin{equation} \label{eq:volphidef}
\hat \nu_{|\phi_o} \, \tpsi(\nu,\phi) = U(\phi - \phi_o) \, |\hat \nu| \tpsi(\nu, \phi_o) 
\end{equation}
and 
\begin{equation}
  \hat p_\phi \, \tpsi(\nu,\phi) = \hbar \, \sqrt{\Theta} \, \tpsi(\nu,\phi) ~.
\end{equation}


In order to obtain the inner product we use the group averaging procedure
\cite{almmt:groupavg,Marolf:groupavg}.  Here one finds a rigging map $\eta:
\Omega \rightarrow \Omega^*$ where $\Omega$ is a dense subspace of the 
(symmetric sector of the) 
auxiliary Hilbert space $L^2(\IR^2, \nu^{-1}d\nu \, d \phi)$.  The states
$(\tpsi| \in \Omega^*$ are defined by considering the following action of the
self-adjoint quantum constraint operator $\hat {H}$:
\begin{equation}
(\tpsi| = \int d \zeta \langle e^{- i \zeta \hat {H}} \tchi |
\end{equation}
for $|\tchi\rangle \in \Omega$.  The inner product then results from the
action $(\tpsi|\tchi\rangle$. It takes on the Schr\"{o}dinger-like form
\cite{aps:improved}
\begin{equation}
\bracket{\Phi}{\Psi} =
\int_{\phi=\phi_o} \frac{d \nu}{\nu} \, \bar \Phi(\nu,\phi) \, \Psi(\nu,\phi) ~
\end{equation}
in the $(\nu,\phi)$ representation,%
\footnote{The form of the inner product depends on the choice of
representation.  In terms of the curvature variable
$y=\ln(\beta/\beta_o)/\cons$, for example -- restricting to $\beta \geq 0$ for
the same reason we restrict to $\nu \geq 0$ -- the inner product takes on the
Klein-Gordon form $\bracket{\Phi}{\Psi}=-i\int_{\phi=\phi_o}dy\
\bar{\Phi}\stackrel{\leftrightarrow}{\partial}_{\phi}\Psi$ \cite{acs:slqc}.
\label{fn:ipform}
} %
and under it the action of the Dirac observables is self-adjoint.   (Here and 
henceforth all $\nu$ integrations will be taken over the range $0 \leq \nu < 
\infty$ unless otherwise specified.  Since the states are symmetric in $\nu$, 
the only differences this choice induces are absolute value signs on the 
factor of $\nu$ in the measure, and $\sqrt{2}$'s in normalization.)

In this inner product the left- and right-moving subspaces (respectively,
contracting and expanding in the volume representation) are orthogonal, and
matrix elements between them of any operator which leaves these subspaces
invariant will be zero.

Equipped with the essential apparatus of the quantum theory we may now proceed
to define projection operators for the volume observable in the physical
Hilbert space.  First note that the eigenfunctions of
$\Theta(\nu)$ satisfy the identity
\begin{equation}\label{eq:eigen_nu_ortho}
\int_{-\infty}^\infty d k \, \bar e_k(\nu') \, e_k(\nu) = 
\delta(\ln \nu - \ln \nu') ~.
\end{equation}
Using the resolution of the identity in $k$
\begin{equation}
\int_{-\infty}^\infty d k |k\rangle \langle k| = 1,
\end{equation}
we can rewrite Eq.(\ref{eq:eigen_nu_ortho}) in bra-ket notation (with 
$e_k(\nu)=\bracket{\nu}{k}$) as
\begin{equation}
\langle \nu|\nu' \rangle = \delta(\ln  \nu  - \ln \nu') ~.
\end{equation}
The resolution of the identity for volume can then be written as%
\footnote{Note that the factor $\nu$ in the denominator is required by
consistency with our normalization of the states and the inner product.  With
a different choice of normalization of states it is possible to rewrite the
identity in a form in which $\nu$ does not appear in the denominator.
} %
\begin{equation}
\int_0^\infty \frac{d \nu}{\nu} \, |\nu \rangle \langle \nu| = 1 ~.
\label{eq:volident}
\end{equation}
We can now define the projector for the volume in the range $\Delta \nu$.
This is obtained from Eq.\ (\ref{eq:volident}) by restricting the integration
to $\Delta \nu$:
\begin{equation}\label{eq:projvol}
P_{\Delta\nu} = 
\int_{\Delta\nu} \! \frac{d \nu}{\nu} \, |\nu \rangle \langle \nu| 
= \int_{\Delta\nu} \! d P_{\nu} ~,
\end{equation}
where $d P_{\nu}$ is an infinitesimal projector.  In the next section we will
see that given this projector one can consistently define class operators for
the histories in which volume lies in the range $\Delta \nu$, leading to a
natural consistent probability interpretation for the volume observable in
agreement with the one obtained from the action of the corresponding Dirac
observable.

\subsection{Semiclassical States}
\label{sec:scstates}

Of particular interest will be quantum states which are semiclassical in the 
sense of being peaked on classical trajectories, Eq.\ (\ref{eq:cltraj}).  An 
example is the state
\begin{equation}
\Psi_{\mathit{sc}}(\nu,\phi) = 
\frac{1}{\sqrt{2\pi}}\frac{1}{\sqrt{\sqrt{\pi}\sigma}}
   \int_{-\infty}^\infty \, d k \, 
   e^{-(k - \bar{k})^2/2 \sigma^2}  e^{ik\ln\nu}
   e^{i \omega(\phi - \bar{\phi}_{\mp})},
\label{eq:semiclass}
\end{equation}
where $\omega = \sqrt{12\pi G}|k|$ and $\bar{\phi}_{\pm}=\pm
\ln\bar{\nu}/\sqrt{12\pi G} + \phi_o$, with the top sign for
expanding/right-moving ($\bar{k}<0$) and the bottom for
contracting/left-moving ($\bar{k}>0$) solutions.  If $\bar{\nu}\gg 1$
(equivalently, $\bar{V}\gg l_p^3$) and $\bar{p}_{\phi} =\sqrt{12\pi G}\hbar
|\bar{k}| \gg\hbar$ at $\phi=\phi_o$, this state remains peaked on the
classical solution specified by $\nu_o=\bar{\nu}$ at $\phi=\phi_o$ throughout
its evolution \cite{acs:slqc,cor-singh08a}.
For the flat FRW model with a massless scalar field this amounts to specifying
semi-classical initial conditions at small spacetime curvatures in a large
universe.

Indeed, for this state it is straightforward to verify that while
\begin{subequations}
\begin{eqnarray}
\sigma_{\ln\nu} & = &
 \sqrt{\expct{(\ln\nu)^2} -\expct{\ln\nu}^2} 
\label{eq:lnnuspread-a}\\
&=& \frac{1}{\sqrt{2}\sigma},
\label{eq:lnnuspread-b}
\end{eqnarray}
\label{eq:lnnuspread}%
\end{subequations}
the value of $\sigma_{\nu} = \sqrt{\expct{\nu^2}-\expct{\nu}^2}$ grows 
exponentially away from the singularity on both branches:
\begin{equation}
\sigma_{\nu}^2 = 
e^{\pm2(\phi-\bar{\phi}_{\pm})} e^{\frac{1}{2\sigma^2}}(e^{\frac{1}{2\sigma^2}} -1).
\label{eq:nuspread}
\end{equation}
(See Sec.\ \ref{sec:volphi} for discussion of how these expectation values
are calculated.)

\section{The decoherence functional}
\label{sec:dWdW}

In this section, we describe the elements necessary to construct the
decoherence functional for the model described in section \ref{sec:WdWmodel}.
In Ref.\ \cite{CH04}, a decoherence functional was constructed for a path integral
quantization of (closed) type A minisuperspace models.  Our definitions will
largely follow the framework established in that work, with modifications
appropriate to an open universe and canonical quantization.

We describe in sequence the required elements: the class operators which
describe physical histories (Section \ref{sec:classops}); the corresponding
branch wave functions, the amplitudes for each such history for a given state
(Section \ref{sec:bwfs}); and finally, the decoherence functional which
measures the interference among the branch wave functions, and, when that
interference vanishes, their probabilities (Section \ref{sec:df}).  We will
see that, with suitable adaptions of the basic definitions, the construction
closely resembles that of ordinary non-relativistic particle mechanics.

\subsection{Class Operators}
\label{sec:classops}

Class operators describe potential physical histories -- most generally, 
coarse-grained histories such as defined in Eq.\ (\ref{eq:classopcgqm}).
Homogeneous class operators describe possible sequences of (ranges of) values
of observable quantities, with sums of them corresponding to coarse-grainings
thereof.   We will often refer to class operators simply as ``histories''.

Class operators correspond to the physical questions that may be asked
of a given system.  For the model under consideration these include
``What is the physical volume of the fiducial cell when the scalar
field has value $\phi^*$?''  ``Does the volume of the cell ever drop
below a particular value, let us say $\nu^*$?''  ``Is the momentum of
the scalar field conserved during evolution?'' 
-- and so forth.  All such questions come in exclusive, exhaustive sets -- at
the most coarse-grained level, simply ``Does the universe have property
${\mathcal P}$, or not?''  The sum of all the class operators in such an
exclusive, exhaustive set must therefore be, in effect, the identity -- as
expressed in Eq.\ (\ref{eq:classopsumqm}) -- up to a possible overall unitary
factor (see Section \ref{sec:bwfsolns}).

In non-relativistic quantum theory fine-grained class operators are given by 
Eq.\ (\ref{eq:classopdefqm}).  In quantum cosmology they may be constructed 
similarly for many questions of physical interest.

In the model at hand, we have states $\ket{\Psi}$ with a unitary evolution in
$\phi$ given by Eq.\ (\ref{eq:WdWprop}).  Physical quantities of interest will
include the values of volume and scalar momentum at given values of $\phi$.
To extract physical predictions concerning quantities of this kind, let us
proceed as in ordinary quantum theory and define ``Heisenberg projections''
\begin{equation}
\Projsupb{\alpha}{\Delta a^{\alpha}_k}(\phi) = 
  U^{\dagger}(\phi-\phi_0) \Projsupb{\alpha}{\Delta a^{\alpha}_k} U(\phi-\phi_0),
\label{eq:HprojDqc}
\end{equation}
where $\Projsupb{\alpha}{\Delta a^{\alpha}_k}$ is the projection onto the 
range of eigenvalues $\Delta a^{\alpha}_k$ of the operator $A^{\alpha}$
and $\phi_0$ is a fiducial value of the scalar field at which the quantum 
state is defined.%
\footnote{Because the evolution in $\phi$ is unitary this fiducial value is 
completely arbitrary and may be adjusted at will to remain outside the region 
of coarse-grainings of physical interest.
} %
The coarse-grained history
\begin{equation}
h = 
(\Delta a^{\alpha_1}_{k_1},\Delta a^{\alpha_2}_{k_2},\dots,\Delta a^{\alpha_n}_{k_n})
\label{eq:historyqc}
\end{equation}
in which the variable $\alpha_1$ takes values in $\Delta a^{\alpha_1}_{k_1}$ 
at $\phi=\phi_1$, variable $\alpha_2$ takes values in $\Delta a^{\alpha_2}_{k_2}$ 
at $\phi=\phi_2$, and so on, then has the class operator
\begin{subequations}
\begin{eqnarray}
C_h &=& \Projsupb{\alpha_1}{\Delta a_{k_1}}(\phi_1)
        \Projsupb{\alpha_2}{\Delta a_{k_2}}(\phi_2) \cdots
        \Projsupb{\alpha_n}{\Delta a_{k_n}}(\phi_n) 
\label{eq:classopdefqc-a}  \\
    &=& U(\phi_0-\phi_1)\Projsupb{\alpha_1}{\Delta a_{k_1}}
        U(\phi_1-\phi_2)\Projsupb{\alpha_2}{\Delta a_{k_2}} \cdots
        U(\phi_{n-1}-\phi_n)\Projsupb{\alpha_n}{\Delta a_{k_n}}U(\phi_n-\phi_0), 
\label{eq:classopdefqc-c}
\end{eqnarray}
\label{eq:classopdefqc}%
\end{subequations}
where again we suppress the superscripts on the eigenvalue ranges to minimize 
the clutter.  They are normalized according to 
\begin{subequations}
\begin{eqnarray}
\sum_h C_h & = & \sum_{k_1}\sum_{k_2}\cdots \sum_{k_n} \, 
                     \Projsupb{\alpha_1}{\Delta a_{k_1}}(\phi_1)
                     \Projsupb{\alpha_2}{\Delta a_{k_2}}(\phi_2) \cdots
                     \Projsupb{\alpha_n}{\Delta a_{k_n}}(\phi_n)
\label{eq:classopsumqc-a}\\
&=& \sum_{k_1}\sum_{k_2}\cdots \sum_{k_n} \,  U(\phi_0-\phi_1) \Projsupb{\alpha_1}{\Delta a_{k_1}} 
U(\phi_1-\phi_2) \cdots U(\phi_{n-1}-\phi_n) \Projsupb{\alpha_n}{\Delta a_{k_n}}  U(\phi_n-\phi_0) \nonumber \\ \\
& = & U(\phi_0-\phi_1)  U(\phi_1-\phi_2) \cdots U(\phi_{n-1}-\phi_n) U(\phi_n-\phi_0)
\label{eq:classopsumqc-b}\\
& = & U(\phi_0-\phi_n) U(\phi_n-\phi_0)
\label{eq:classopsumqc-c}\\
 & = & 1, 
\label{eq:classopsumqc-d}
\end{eqnarray}
\label{eq:classopsumqc}%
\end{subequations}
the identity on the physical Hilbert space $\Hphys$.

For example, the class operator for the history in which the volume $\nu$ is 
in $\Delta\nu$ when the scalar field $\phi=\phi^*$ is simply
\begin{equation}
C_{\Delta\nu|_{\phi^*}}  =  
  U^{\dagger}(\phi^*-\phi_0) \Projsupb{\nu}{\Delta \nu} U(\phi^*-\phi_0),
\label{eq:classopvol}
\end{equation}
where $\Projsupb{\nu}{\Delta \nu}$ is given by Eq.\ (\ref{eq:projvol}).  Note
that we employ here projections onto ranges of values of the volume operator
$\hat{\nu}$, \textbf{not} the Dirac observable $\hat{\nu}|_{\phi^*}$.  We will
return to this point in Section \ref{sec:relational}, once we have completed
the definition of the decoherence functional.  However, comparison of Eqs.\
(\ref{eq:classopvol}) and (\ref{eq:volphidef}) or (\ref{eq:relnlobsA}) should
be sufficient to suggest the thrust of that discussion.


\emph{Remark:} It is certainly not the case that \emph{all} questions of
physical interest may be directly captured by class operators of the specific
form defined by Eq.\ (\ref{eq:classopdefqc}) or coarse-grainings thereof.
Indeed, this is not the case even in ordinary quantum theory.  We give two
examples.

First, consider physical questions defined by coarse-grainings of
minisuperspace by two-dimensional regions \emph{i.e.\ }coarse-grainings by
continuous ranges of $\phi$ and other variables.  In the current quantization,
where $\phi$ plays in effect the mathematical role of an emergent ``time'',
such questions are akin to questions in particle mechanics such as ``Does the
particle ever enter spacetime region $R$?''  There are various approaches for
tackling this and related problems, both in ordinary quantum mechanics
\cite{yt91a,*yt91b,*yt92,y92,*y96,lesH,whelan94,micanek96} and in quantum cosmology
\cite{hartlemarolf97,hallithor01,hallithor02,CH04,halliwall06,as05,halliwell09},
but we shall not take up questions of this kind here -- not least, because
with currently available methods, it is very difficult to calculate anything
except in the very simplest of cases even in particle mechanics.%
\footnote{Apropos of this point, it may be worth noting that it should be
possible to express the decoherence functional in a form which is manifestly
diffeomorphism invariant; see \cite{CH04} for an example of what this looks
like in a different (functional integral) quantization, a general discussion
of diffeomorphism invariant physical alternatives, and one way of constructing
the corresponding class operators.  
Alternative approaches which demand the class operators commute with the
constraint are developed in
\cite{hartlemarolf97,hallithor01,hallithor02,halliwall06,as05,halliwell09}.
In the present canonical context, note that instead our \emph{branch wave
functions} -- \emph{cf.\ }Eq.~(\ref{eq:bwfdefqc}) -- are solutions of the
constraint; see Sec.\ \ref{sec:bwfsolns}.
} %

As a second example, we can consider a reformulation of the present model in
which the scale factor (or volume) is treated as a clock rather than the
scalar field.  The simplicity of the model allows this unitary equivalence
between ``space'' (labeled by volume) and ``time'' (labeled by the field).  In
this way questions such as ``What is the value of the scalar field at $\nu =
\nu_o$?''  may be coherently posed.  Regardless, we will not take up questions
of this kind here.

\subsection{Branch wave functions}
\label{sec:bwfs}

Class operators, as described in the previous sub-section, capture the
physical questions that may be asked of a system.  The branch wave functions
$\ket{\Psi_h}$ constructed from an exclusive, exhaustive set of histories $\{
h \}$, and a given quantum (``initial'') state $\ket{\Psi}$ specified at
$\phi=\phi_o$, are then the amplitudes for that state to ``follow'' the
histories $h$ \emph{i.e.\ }for the universe to have the properties described
by $h$.

Branch wave functions in non-relativistic particle mechanics are defined by
Eq.\ (\ref{eq:bwfdefqm}), or Eq.\ (\ref{eq:bwfdefqmt}) to define a state on
spacetime that satisfies the Schr\"{o}dinger equation everywhere.  In quantum
cosmology, the branch wave function for a state $\ket{\Psi}$ (in the physical
Hilbert space) and history $h$ may be constructed in the same way:
\begin{equation}
\ket{\Psi_h(\phi)} = U(\phi-\phi_0)C_h^{\dagger}\ket{\Psi}.
\label{eq:bwfdefqc}
\end{equation}
This branch wave function is, by construction, a solution to the 
Wheeler-DeWitt equation everywhere.
The extra propagator $U$, of course, simply evolves the branch wave function
to any convenient $\phi=\text{constant}$ slice.  (All inner products will of
course be independent of this choice.)


\subsection{Decoherence functional}
\label{sec:df}

Given a complete exclusive, exhaustive set of histories $\{ h \}$ and a
quantum state $\ket{\Psi}$, the decoherence functional measures the
interference among the branch wave functions $\ket{\Psi_h}$, and, if that
interference vanishes, determines also the probabilities of each of the
$\ket{\Psi_h}$ -- in other words, the probability that a universe in the state
$\ket{\Psi}$ has the properties described by the history $h$.  If the
interference does not vanish, then the set of physical questions $\{ h \}$
does not make sense in the quantum theory, in exactly the same way the
question of which slit a particle passed through cannot be coherently analyzed
when it is not recorded.

The decoherence functional in non-relativistic quantum mechanics is defined 
according to Eq.\ (\ref{eq:dfdefqm}).  In quantum cosmology the decoherence 
functional may be constructed from the branch wave functions in essentially 
the same manner \cite{CH04}:%
\footnote{The (reasonably transparent) generalization of this definition to 
mixed states is also given in \cite{CH04}.
} %
\begin{equation}
d(h,h') =  \bracket{\Psi_{h'}}{\Psi_{h}}.
\label{eq:dfdefqc}\\
\end{equation}
Decoherent sets of histories satisfy
\begin{equation}
d(h,h') = p(h)\, \delta_{h'h},
\label{eq:ffqc}
\end{equation}
where $p(h)$ is the probability for the history $h$.
Note that as constructed, the decoherence functional for this model involves 
an inner product of branch wave functions on a minusuperspace slice of fixed 
$\phi$.  
The unitary evolution in $\phi$, and the fact that the branch wave functions
$\ket{\Psi_h(\phi)}$ are by construction everywhere solutions of the
Wheeler-DeWitt equation, makes the specific choice of $\phi$ irrelevant in the
definition of the branch wave functions and decoherence functional, and
therefore may be selected for maximal convenience.  This freedom will be
exploited frequently in the sequel.

In more general models where a unitary evolution in some effective clock
variable like $\phi$ might not be available, it is more natural to define Eq.\
(\ref{eq:dfdefqc}) on (mini)superspace slices that are \emph{spacelike} in the
DeWitt metric, \emph{e.g.\ }on slices of constant scale factor -- as is done,
for example, in Ref.\ \cite{CH04}.  In this simple model, however, these
choices are unitarily equivalent, as should be evident from Eq.\
(\ref{eq:wdw}).

Indeed, so long as we can define normalizeable states on slices of superspace
that are spacelike in the DeWitt metric, and a unitary evolution between such
slices, we might expect Eq.\ (\ref{eq:dfdefqc}) -- or one of its simple
generalizations, as in Ref.\ \cite{CH04} -- to be a suitable definition for
the decoherence functional in a full theory of quantum gravity as well.

\subsection{Histories and relational observables}
\label{sec:relational}

We noted below Eq.\ (\ref{eq:classopvol}) that the class operator for volume
was defined using projections onto ranges of the volume operator $\hat{\nu}$,
which is \emph{not} a Dirac observable.  We now clarify the relation between
probabilities computed from class operators and the Dirac observables.

To see this, consider a self-adjoint local operator $\hat{A}$ that does not
commute with the constraint and the corresponding relational observable given
by Eq.\ (\ref{eq:relnlobsA}).  Assuming for definiteness the spectrum of
$\hat{A}$ to be continuous, take formally the spectral resolution of $\hat{A}$
to be
\begin{subequations}
\begin{eqnarray}
\hat{A} & = & \int\! da\ a\, \ketbra{a}{a}
\label{eq:projA-a}\\
& = & \int\! da\ a\, P_a
\label{eq:projA-b}\\
& = &  \int\! a\; dP_a,
\label{eq:projA-c}
\end{eqnarray}
\label{eq:projA}%
\end{subequations}
with $P_a\equiv\ketbra{a}{a}$ and $dP_a\equiv P_a\, da$.  

Class operators for ranges of values $\Delta a$ of $\hat{A}$ may be
constructed as
\begin{equation}
C_{\Delta a|_{\phi^*}}  =  
  U^{\dagger}(\phi^*-\phi_0) \Projsupb{A}{\Delta a} U(\phi^*-\phi_0),
\label{eq:classopA}
\end{equation}
the corresponding branch wave functions for a given state $\ket{\Psi}$ being
given by Eq.\ (\ref{eq:bwfdefqc}),
\begin{equation}
\ket{\Psi_{\Delta a|_{\phi^*}}(\phi)} = 
  U(\phi-\phi_0) C_{\Delta a|_{\phi^*}}^{\dagger} \ket{\Psi}.
\label{eq:bwfA}
\end{equation}
Since the class operators are simply projections, the corresponding histories
always decohere.  (See Section \ref{sec:volphi} for details of the calculation
in the case of the volume operator.)  The probability that $a\in\Delta a$ is
then given by
\begin{subequations}
\begin{eqnarray}
p_{\Delta a}(\phi^*) & = & 
\bracket{\Psi_{\Delta a|_{\phi^*}}}{\Psi_{\Delta a|_{\phi^*}}}
\label{eq:probA-a}\\
& = & 
\melt{\Psi}{C_{\Delta a|_{\phi^*}}C_{\Delta a|_{\phi^*}}^{\dagger}}{\Psi}
\label{eq:probA-b}\\
& = & 
\melt{\Psi}{C_{\Delta a|_{\phi^*}}}{\Psi}
\label{eq:probA-c}\\
& = & 
\melt{\Psi}{U^{\dagger}(\phi^*-\phi_0) \Projsupb{A}{\Delta a} U(\phi^*-\phi_0)}{\Psi}
\label{eq:probA-d}\\
& = & 
\melt{U(\phi^*-\phi_0) \Psi}{\Projsupb{A}{\Delta a} U(\phi^*-\phi_0)}{\Psi}
\label{eq:probA-e}\\
& = & 
\melt{\Psi(\phi^*)}{\Projsupb{A}{\Delta a}}{\Psi(\phi^*)}.
\label{eq:probA-f}
\end{eqnarray}
\label{eq:probA}%
\end{subequations}
Taking $\Delta a$ to be the infinitesimal interval $da$, the probability that 
$a\in da$ at $\phi=\phi^*$ is
\begin{equation}
dp_a(\phi^*) = \melt{\Psi(\phi^*)}{dP_a}{\Psi(\phi^*)}.
\label{eq:probAdp}
\end{equation}

To see the connection with the relational observable Eq.\ (\ref{eq:relnlobsA}),
let us find the average value of $\hat{A}$ at $\phi=\phi^*$.   If $\hat{A}$ 
has a discrete spectrum this is simply $\sum_a a\, p_a(\phi^*)$.   Since we 
have taken $\hat{A}$ to have a continuous spectrum,
\begin{subequations}
\begin{eqnarray}
\expct{\hat{A}}|_{\phi^*} & = & \int\! a\, dp_a(\phi^*)
\label{eq:expectrlnl-b}\\
& = & \int\! a\, \melt{\Psi(\phi^*)}{dP_a}{\Psi(\phi^*)}
\label{eq:expectrlnl-c}\\
& = &  \melt{\Psi(\phi^*)}{\int\! a\, dP_a}{\Psi(\phi^*)}
\label{eq:expectrlnl-d}\\
& = & \melt{\Psi(\phi^*)}{\hat{A}}{\Psi(\phi^*)}
\label{eq:expectrlnl-e}\\
& = & \melt{U(\phi^*-\phi_0)\Psi}{\hat{A}\, U(\phi^*-\phi_0)}{\Psi}
\label{eq:expectrlnl-f}\\
& = & \melt{\Psi}{U(\phi^*-\phi_0)^{\dagger}\hat{A}\, U(\phi^*-\phi_0)}{\Psi}
\label{eq:expectrlnl-g}\\
& = & \melt{\Psi}{\hat{A}|_{\phi^*}}{\Psi},
\label{eq:expectrlnl-h}
\end{eqnarray}
\label{eq:expectrlnl}%
\end{subequations}
where the last step follows by comparison with Eq.\ (\ref{eq:relnlobsA}). 
(The calculation is essentially unchanged if the spectrum of $\hat A$
is chosen to be discrete.)\  
We can thus see that the average value of $\hat{A}$ at $\phi=\phi^*$ is
naturally given by the expectation value of the relational observable
$\hat{A}|_{\phi^*}$ in the state $\ket{\Psi}$.  In other words,
\emph{probabilities for histories of values of $\hat{A}$, which does not
commute with the constraint, are naturally expressed in terms of the
corresponding Dirac observable $\hat{A}|_{\phi^*}$, 
which does.}
Put another way, histories formulations of quantum theory provide a natural
framework for understanding the emergence of relational Dirac observables in
theories with constraints.%
\footnote{Note that it is not the role of $\phi$ as an effective ``time'' that
is relevant here.  Rather, what is important is the scheme for predictions
concerning \emph{correlations} between the values of different quantum
variables.
} %


We hope to explore this connection between histories formulations and
relational observables further in another work.

\section{Applications}
\label{sec:app}

Now that the machinery for a consistent histories formulation of quantum 
cosmology in minisuperspace has been fully defined, we turn to application of 
the theory to extract physical predictions.  
We shall examine in turn predictions concerning the scalar momentum, the 
volume of the universe, 
the semiclassical behaviour of the universe, and, finally, the question of
whether a quantum universe shares the inevitably singular fate of its
classical counterpart.

In each case the methodology for prediction is the same.  For each physical 
question the corresponding class operators for an exclusive, exhaustive 
partition of the possible histories must be determined.  Given the class 
operators and a choice of quantum state, it may be ascertained whether or not 
the set of histories decoheres.  For some classes of questions, decoherence 
(or lack thereof) is generic, independent of the choice of state.  In 
general, however, decoherence depends on the state.

If the family of histories fails to decohere, then the question as formulated
does not make sense in the quantum theory, in that probabilities may not be
consistently assigned to the alternative histories.  If the family does
decohere, then probabilities may be assigned according to Eq.\
(\ref{eq:ffqc}), and the relative likelihood of the alternatives assessed.

We now proceed with this plan 
for each of our observable quantities.

\subsection{Volume}
\label{sec:volume}

We will examine predictions concerning the volume of (a fiducial cell of) a
quantum universe, focusing on the specific cases of the volume at a given
value of the scalar field, and of the volume at a \emph{sequence} of values of
the scalar field.  In the first instance we shall find that decoherence
is automatic, essentially because it is a prediction concerning the value of a
single quantity on a single slice.  For given choices of state we show how to
calculate the probabilities explicitly, and exhibit a few examples.  

In the second instance, sequences of values of the volume at different values
of the scalar field, decoherence is considerably more intricate, and indeed,
in general will not occur.
In subsequent sections we will exhibit two examples for which such histories
\emph{do} decohere, and again show how to calculate the probabilities and
illustrate with some examples.

We will employ these results to study in Section \ref{sec:singular} the
question of whether our model quantum universes are singular in an appropriate
sense.


\subsubsection{Volume at a given value of $\phi$}
\label{sec:volphi}

The class operators for the question, ``What is the volume of (a fiducial cell 
of) the universe when $\phi=\phi^*$?'' are given by Eq.\ 
(\ref{eq:classopvol}):
\begin{subequations}
\begin{eqnarray}
C_{\Delta\nu|_{\phi^*}} & =  &
  U^{\dagger}(\phi^*-\phi_0) \Projsupb{\nu}{\Delta \nu} U(\phi^*-\phi_0),
\label{eq:classopvolredux-a}\\
& = &  \Projsupb{\nu}{\Delta \nu}(\phi^*), \label{eq:classopvolredux-b}
\end{eqnarray}
\label{eq:classopvolredux}%
\end{subequations}
where the ranges $\Delta\nu$ are chosen from a set $\{\Delta\nu_i\}$ of
disjoint intervals that partition the full range of volumes $0\leq\nu<\infty$,
so that
\begin{equation}
\sum_{i} C_{\Delta\nu_i|_{\phi^*}} = 1,
\label{eq:classopvolsum}
\end{equation}
the identity on $\Hphys$.  The corresponding branch wave functions are 
\begin{equation}
\ket{\Psi_{\Delta\nu|_{\phi^*}}(\phi)} = 
  U(\phi-\phi_0) C_{\Delta\nu|_{\phi^*}}^{\dagger} \ket{\Psi},
\label{eq:bwfvol}
\end{equation}
where $\ket{\Psi}$ is the quantum state of the universe defined on a chosen
slice $\phi=\phi_0$.

Since $C_{\Delta\nu|_{\phi^*}}$ is simply a projection, it is clear that the
$\{C_{\Delta\nu_i|_{\phi^*}}\}$ are orthogonal when the ranges are different:
\begin{subequations}
\begin{eqnarray}
C_{\Delta\nu_i|_{\phi^*}}^{\phantom{\dagger}} \cdot C_{\Delta\nu_j|_{\phi^*}}^{\dagger} & = & 
  C_{\Delta\nu_i\cap\Delta\nu_j|_{\phi^*}}
\label{eq:classopprodvol-a}\\
& = & 
C_{\Delta\nu_i|_{\phi^*}} \cdot\, \delta_{ij},
\label{eq:classopprodvol-b}\\
& = & 
C_{\Delta\nu_i|_{\phi^*}}^{\phantom{\dagger}} C_{\Delta\nu_i|_{\phi^*}}^{\dagger} \cdot\, \delta_{ij},
\label{eq:classopprodvol-c}
\end{eqnarray}
\label{eq:classopprodvol}%
\end{subequations}
from which it is equally clear that
\begin{subequations}
\begin{eqnarray}
\bracket{\Psi_{\Delta\nu_i|_{\phi^*}}}{\Psi_{\Delta\nu_j|_{\phi^*}}}
& = & 
\bracket{C_{\Delta\nu_i|_{\phi^*}}^{\dagger} \Psi}{C_{\Delta\nu_j|_{\phi^*}}^{\dagger} \Psi}
\label{eq:decohvol-a}\\
& = & 
\bracket{\Psi}{C_{\Delta\nu_i|_{\phi^*}}^{\phantom{\dagger}}  C_{\Delta\nu_j|_{\phi^*}}^{\dagger} \Psi}
\label{eq:decohvol-b}\\
& = & 
\bracket{\Psi_{\Delta\nu_i|_{\phi^*}}}{\Psi_{\Delta\nu_i|_{\phi^*}}}\,\cdot\, \delta_{ij}
\label{eq:decohvol-c}
\end{eqnarray}
\label{eq:decohvol}%
\end{subequations}
for any choice of state $\ket{\Psi}$, and thus this family of histories
decoheres.  From Eq.\ (\ref{eq:classopvolredux-a}), the probability that the
volume of (a fiducial cell of) the universe lies in $\Delta\nu$ when
$\phi=\phi^*$ is then given by
\begin{subequations}
\begin{eqnarray}
p_{\Delta\nu}(\phi^*) & = & 
\bracket{\Psi_{\Delta\nu|_{\phi^*}}}{\Psi_{\Delta\nu|_{\phi^*}}}
\label{eq:probvol-a}\\
& = & 
\melt{\Psi}{C_{\Delta\nu|_{\phi^*}}^{\dagger}}{\Psi}
\label{eq:probvol-b}\\
& = & 
\melt{\Psi}{U^{\dagger}(\phi^*-\phi_0) \Projsupb{\nu}{\Delta \nu} U(\phi^*-\phi_0)}{\Psi}
\label{eq:probvol-c}\\
& = & 
\melt{U(\phi^*-\phi_0) \Psi}{\Projsupb{\nu}{\Delta \nu} U(\phi^*-\phi_0)}{\Psi}
\label{eq:probvol-d}\\
& = & 
\melt{\Psi(\phi^*)}{\Projsupb{\nu}{\Delta \nu}}{\Psi(\phi^*)}
\label{eq:probvol-e}\\
& = & 
\int_{\Delta\nu}\frac{d\nu}{\nu}|\Psi(\nu,\phi^*)|^2
\label{eq:probvol-f}
\end{eqnarray}
\label{eq:probvol}%
\end{subequations}
using Eq.\ (\ref{eq:projvol}).  While this indeed may have been the expected
result, it is important to note that it has here been 
\emph{derived} within a fully coherent framework for constructing quantum
probabilities.  It is also worth emphasizing that the form of the result is
crucially dependent on the form of the inner product in the volume
representation.  It is essential to know the form of the inner product and
representative projections in a representation before drawing any conclusions
concerning the interpretation of the wave function as a probability density
from a formula like Eq.\ (\ref{eq:probvol-a}).  In other representations the
formula for single probabilities will take on a different form.  For example,
for the curvature variable $y$ noted in footnote \ref{fn:ipform}, the
probability formula for $y$ takes on a Klein-Gordon form.  This clearly
illustrates the profound importance of framing quantum probabilities within a
clear and self-consistent framework.

Note in addition that with $\Delta\nu$ taken to be an infinitesimal interval
$d\nu$, Eq.\ (\ref{eq:probvol-f}) -- in conjunction with Eqs.\
(\ref{eq:projA}) and (\ref{eq:probAdp}) -- can then by the calculation of Eq.\
(\ref{eq:expectrlnl}) be employed to determine, in any given representation,
the corresponding expression for an expectation value. For example,
$\expct{\nu^n}=\int\nu^n dp_{\nu}=\int\frac{d\nu}{\nu}\nu^n|\Psi(\nu,\phi)|^2$.

In Section \ref{sec:singular} we shall apply these results to a discussion of 
whether or not the quantum universe is singular, in the sense that the 
probability that the volume of (a fiducial cell of) the universe becomes zero 
at some value of the scalar field is unity for a given state $\ket{\Psi}$.

We complete this sub-section by calculating these probabilities for the
semiclassical state given in Eq.\ (\ref{eq:semiclass}).  It is straightforward
to show from Eq.\ (\ref{eq:probvol}) that the probability that such a universe
will be found with volume in the range $\Delta\nu=[\nu_1,\nu_2]$ at $\phi$ is
given by
\begin{subequations}
\begin{eqnarray}
 p^{\mathit{sc}}_{\Delta\nu}(\phi)    & = & 
\frac{1}{2}\left\{
\erf(\sigma(\ln\nu_2 \mp \cons(\phi-\bar{\phi}_{\mp}))) -
\erf(\sigma(\ln\nu_1 \mp \cons(\phi-\bar{\phi}_{\mp})))\right\}
\label{eq:pvolsc-a}\\  & = & 
\frac{1}{2}\left\{
\erf(\sigma(\ln\frac{\nu_2}{\bar{\nu}} \mp \cons(\phi-\phi_o))) -
\erf(\sigma(\ln\frac{\nu_1}{\bar{\nu}} \mp \cons(\phi-\phi_o)))\right\}.
\qquad\label{eq:pvolsc-b}
\end{eqnarray}
\label{eq:pvolsc}%
\end{subequations}
Here the upper sign is for expanding solutions and the lower, contracting.
Thus, for example, the likelihood that a semiclassical universe will be found
at small volume -- \emph{i.e.\ }with volume less than $\nu^*$ -- is
\begin{equation}
p^{\mathit{sc}}_{\Delta\nu^*}(\phi) = 
\frac{1}{2}\left\{1+\erf(\sigma(\ln\frac{\nu^*}{\bar{\nu}} \mp \cons(\phi-\phi_o)))\right\},
    \label{eq:pvolsc-sm}
\end{equation}
where $\Delta\nu^*=[0,\nu^*]$.

\subsubsection{Volume at a sequence of values of $\phi$}
\label{sec:volphiseq}

Consider a sequence of values of the scalar field
$(\phi_1,\phi_2,\dots,\phi_n)$.  
The class operators for the question, ``What is the volume of (a
fiducial cell of) the universe when $\phi=\phi_1$, again when 
$\phi=\phi_2$, and so forth through  $\phi=\phi_n$?'' 
are
\begin{equation}
C_{\Delta\nu_1|_{\phi_1};\Delta\nu_2|_{\phi_2};\cdots;\Delta\nu_n|_{\phi_n}}  
= \Projsupb{\nu}{\Delta \nu_1}(\phi_1) \Projsupb{\nu}{\Delta \nu_2}(\phi_2) 
      \cdots \Projsupb{\nu}{\Delta \nu_n}(\phi_n),
\label{eq:classopvolseq}
\end{equation}
where the $\Projsupb{\nu}{\Delta \nu}(\phi)$ are given by Eq.\
(\ref{eq:HprojDqc}) and each of the sets of ranges
$(\{\Delta\nu_1\},\{\Delta\nu_2\},\dots,\{\Delta\nu_n\})$ again partition the
interval $0\leq\nu<\infty$.  The branch wave functions are defined in the
usual way, according to Eq.\ (\ref{eq:bwfdefqc}).

Before any question of probabilities for such histories is addressed, it must
be determined whether these histories decohere.  Since the class operators
Eq.\ (\ref{eq:classopvolseq}) are no longer projections, it is neither obvious
nor trivial that they do.  Indeed, in general, histories of this kind will
\textbf{not} be expected to decohere, no more than the histories asking which
slit a particle passed through in the two-slit experiment decohere unless a
recording apparatus is in place.  We will, however, see three important
examples for which they do, in Sections \ref{sec:pphi},
\ref{sec:semiclassical}, and \ref{sec:singular}.

\subsection{Scalar momentum}
\label{sec:pphi}

Classically, the scalar momentum is a constant of the motion,
$\{p_{\phi},H\}=0$ (\emph{cf.\ }Eq.\ (\ref{eq:classconstraintaphi}).)  In the
quantum theory, $[\hat{p}_{\phi},\Theta]\Psi=0$.  Since $\hat{p}_{\phi}$
commutes with the constraint, it is a constant of the motion in the quantum
theory as well.  Thus, defining the relational observable
$\hat{p}_{\phi}|_{\phi^*}$ giving the value of $\hat{p}_{\phi}$ at
$\phi=\phi^*$, we see from Eq.\ (\ref{eq:relnlobs}) that
$\hat{p}_{\phi}|_{\phi^*} = \hat{p}_{\phi}$.  The meaning of this is that if
we carry through a calculation for the probability $p_{\Delta
p_{\phi}}(\phi^*)$ for $p_{\phi}$ to be found in $\Delta p_{\phi}$ similar to
that which led to Eq.\ (\ref{eq:probvol}), we find that $p_{\Delta
p_{\phi}}(\phi^*) = p_{\Delta p_{\phi}}$ is \emph{independent} of $\phi^*$.

Moreover, if one considers histories of the form 
\begin{equation}
C_{\Delta p_{\phi;1}|_{\phi_1};\Delta p_{\phi;2}|_{\phi_2};\cdots;\Delta p_{\phi;n}|_{\phi_n}}  
= \Projsupb{p_{\phi}}{\Delta p_{\phi;1}}(\phi_1) \Projsupb{p_{\phi}}{\Delta p_{\phi;2}}(\phi_2) 
      \cdots \Projsupb{p_{\phi}}{\Delta p_{\phi;n}}(\phi_n),
\label{eq:classoppphiseq}
\end{equation}
since 
$\Projsupb{p_{\phi}}{\Delta p_{\phi}}(\phi) = \Projsupb{p_{\phi}}{\Delta p_{\phi}}$, %
every one of these class operators is zero except for those for which all the
intervals $\Delta p_{\phi}$ are the same, which then reduce simply to a
projection on that range of $p_{\phi}$.  Thus the corresponding branch wave
functions (for any initial state) decohere, and their probabilities are
constant.  This of course is just an expression of the fact that
$\hat{p}_{\phi}$ is a constant of the motion \emph{i.e.\ }does not change with
evolution in $\phi$.

Constants of the motion in generalized quantum theory are discussed in 
greater depth in Ref.\ \cite{hlm94}.






\subsection{Semiclassical evolution}
\label{sec:semiclassical}

More than one meaning may be assigned to the notion that a universe ``behaves
semiclassically''.  The emergence of quasi-classical behavior of a quantum
system typically involves the decoherence of histories corresponding to
approximately classical trajectories due to correlations with -- typically,
but not necessarily -- microscopic degrees of freedom, leading to
quasi-classical equations of motion for the remainder
\cite{GMH93,giulini,schlosshauer07}.  In the present simple model, there are
few degrees of freedom with which to correlate.  Nonetheless, the model still
displays decoherence for quasi-classical histories in the following way.

\begin{figure}[tbh!]
\includegraphics[scale=0.75]{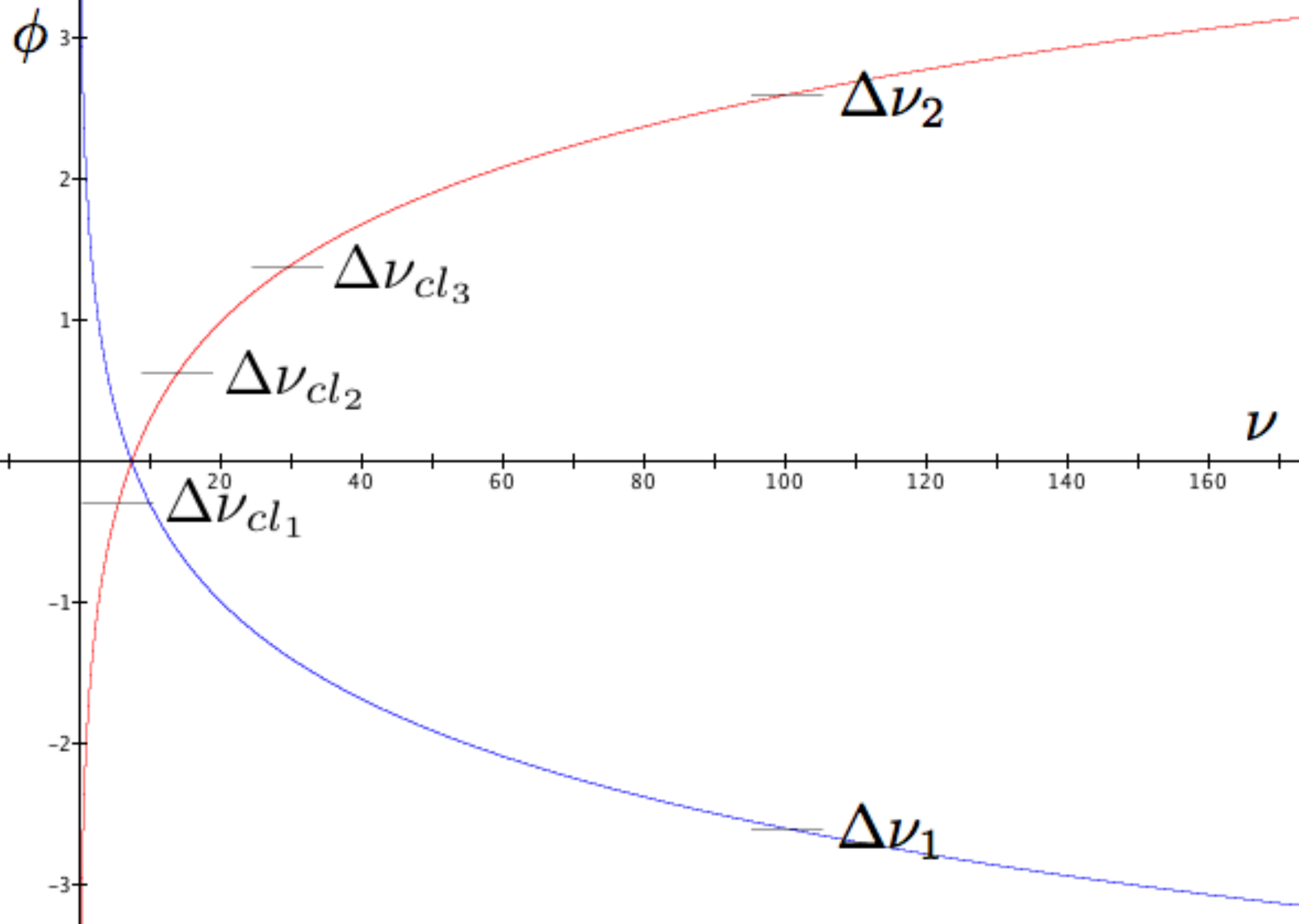}
\caption{Coarse-graining by ranges of values of the volume at different values
of the scalar field.  Histories consisting of ranges which straddle a
particular classical trajectory are the ``(semi-)classical'' histories; all
others describe ``non-classical'' behavior.  Two histories are shown -- one
classical and one not.  The first is a coarse-grained history
$(\Delta\nu_{{cl}_1},\Delta\nu_{{cl}_2},\Delta\nu_{{cl}_3})$ describing a
(semi-)classically expanding universe.  The second history
$(\Delta\nu_1,\Delta\nu_2)$ describes a (very highly coarse-grained) ``quantum
bounce'', which asks whether a quantum universe may be peaked on a classical
solution at both early and late values of $\phi$.  }
\label{fig:classicalcg}
\end{figure}

Consider a semiclassical state (such as Eq.\ (\ref{eq:semiclass})) that is
peaked along some classical trajectory.  Coarse-grain minisuperspace on a set
of slices $\{\phi_1,\phi_2,\dots,\phi_n\}$ by ranges of volume
$\{\Delta\nu_{i_k}, k=1...n\}$ on each slice $\phi_k$
(so that $\cup_{i_k}\Delta\nu_{i_k}=[0,\infty)$ for each $k$),
chosen in such a way that on each slice one of the ranges $\Delta\nu_{{cl}_k}$
straddles the classical trajectory on that slice.%
\footnote{If we continue to regard $\phi$ as an effective ``time'', the
equivalent question in particle mechanics would be, ``Is the particle in
$\Delta x_1$ at time $t_1$, in $\Delta x_2$ at time $t_2$, and so on through
$\Delta x_n$ at time $t_n$?''  Clearly this is a coarse-grained way to inquire
whether the particle has followed a particular path, in this case the one
defined by the intervals $\Delta x_i$.
} %
(See Fig.\
\ref{fig:classicalcg}.)  If the $\Delta\nu_{{cl}_k}$ are chosen to be
comparable in width or wider than%
\footnote{One typical consequence of correlations with other degrees of
freedom in more complicated models is stabilization of semiclassical wave
packets against quantum spreading.
} %
the width given by Eq.\ (\ref{eq:nuspread}), then essentially the only branch
wave function that is not zero is
\begin{equation}
  \ket{\Psi_{cl}} =  \Projsupb{\nu}{\Delta\nu_{{cl}_n}}(\phi_n)  \cdots
     \Projsupb{\nu}{\Delta\nu_{{cl}_2}}(\phi_2)
     \Projsupb{\nu}{\Delta\nu_{{cl}_1}}(\phi_1)\ket{\Psi}. 
\label{eq:qwfclass}
\end{equation}
In this way, the family of histories $(\mathit{classical,nonclassical})$
decoheres, and quasi-classical behavior is predicted for such states with
probability one.  ``Large'' universes are more robustly semiclassical because
their spreading is proportionately small, according to Eq.\
(\ref{eq:nuspread}).
(One can make related arguments for the emergence of appropriately 
quasiclassical behavior for WKB states \cite{hartle91a,lesH,CH04}.)

In models as simple as the present one, this is about as far as one can go.   
For studies of the emergence of quasiclassical behavior in more complex 
systems, see Refs.\ \cite{GMH93,halliwell98,*halliwell99a}.


\subsection{Singularity in a Wheeler-DeWitt quantum universe}
\label{sec:singular}

In quantum theory there are various inequivalent meanings one might assign to 
the idea that a quantum universe is or is not ``singular''.  
In this section we address the question of whether our model quantum universe
is singular with explicit criteria based on the probabilities that various
physical quantities assume values that signal a physical singularity.  We
study in particular the volume observable.  Given that the scalar momentum
$p_\phi$ is a constant of motion in the quantum theory, knowledge concerning
the volume observable directly implies the same for the energy density
$\hat\rho_{| \phi} = (1/2) \hat V^{-1}_{|\phi} \hat p_\phi  \hat V^{-1}_{|\phi}$, %
and hence the corresponding operators for spacetime curvature invariants.
Indeed, the statement that the universe has zero volume in this model is
equivalent to the statement that energy density and spacetime curvature
invariants diverge.  The conclusion will be that, according to \emph{all} of
these criteria, these Wheeler-DeWitt universes are singular for all choices of
state.  This includes, in particular, ``Schr\"{o}dinger's Cat'' states,
generic (but possibly macroscopic) superpositions of left-moving (contracting)
and right-moving (expanding) states.  We will show analytically that in this
quantization, completely generic quantum states of the universe are always
singular.

\subsubsection{Volume singularity}
\label{sec:singularvol}

First, we study the question of the singularity of the universe by inquiring
after the likelihood that the volume of (a fiducial cell of) the universe
becomes zero.  We shall show rigorously that the probability that the universe
has zero volume at some value of the scalar field is unity for all choices of
state.  In particular, we will examine the behaviour of these probabilities
analytically for generic states in the limit that the absolute value of the
scalar field becomes large, with the result that the volume goes to zero with
certainty for all left-moving (contracting) states as $\phi\rightarrow
+\infty$, and for all right-moving (expanding) states as $\phi\rightarrow
-\infty$.  This turn out to be sufficient to imply that the probability that
these universes assume zero volume at \emph{some} value of $\phi$ is unity,
independent of the choice of state $\ket{\Psi}$.

\begin{figure}[tbh!]
\includegraphics[scale=0.75]{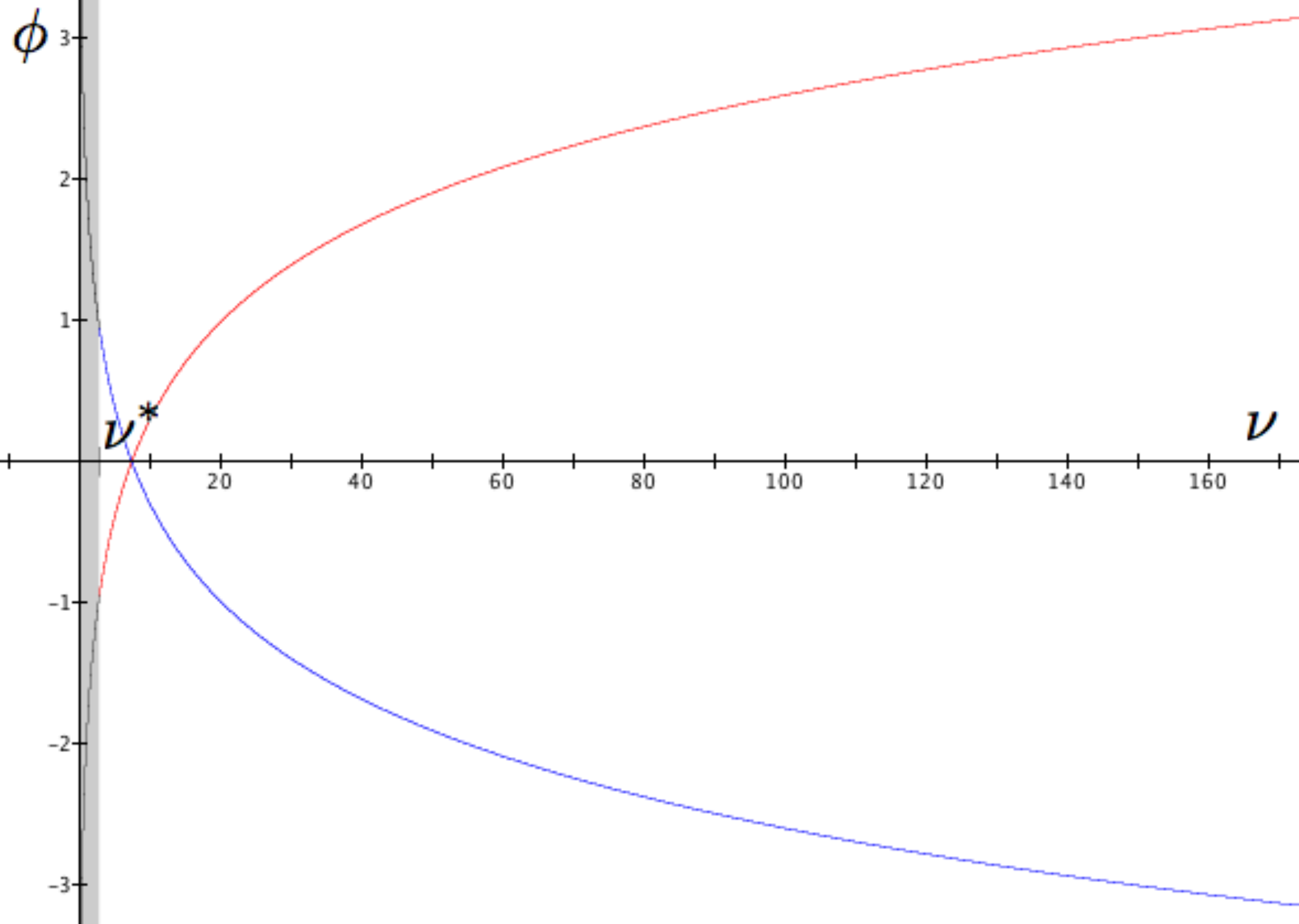}
\caption{Coarse-graining of minisuperspace suitable for studying the
probability that the universe assumes zero volume.  Partition the volume $\nu$
into the range $\Delta\nu^*=[0,\nu^*]$ (the shaded region in the figure) and
its complement $\overline{\Delta\nu^*}=(\nu^*,\infty)$.  The quantum universe
may be said to attain zero volume if the probability for the branch wave
function $\ket{\Psi_{\smash{\Delta\nu^*}}}$ is near unity while that for
$\ket{\Psi_{\smash{\overline{\Delta\nu^*}}}}$ is near zero for arbitrary
choices of $\nu^*$.  }
\label{fig:smallvol}
\end{figure}

To ask the question whether the volume of (a fiducial cell of) the universe
becomes zero, at any value of $\phi$ partition the volume $\nu$ into the range
$\Delta\nu^*=[0,\nu^*]$ for some fixed volume $\nu^*$, and its complement
$\overline{\Delta\nu^*}=(\nu^*,\infty)$.  (See Fig.\ \ref{fig:smallvol}.)
Since this is a coarse-graining defined on a slice of fixed $\phi$, we know
that the histories decohere.  We would like to calculate the probabilities
$p_{\smash{\Delta\nu^*}}(\phi)$ and
$p_{\smash{\overline{\Delta\nu^*}}}(\phi)$.  These will be given by Eq.\
(\ref{eq:probvol}).

To begin, let us consider separately states $\ket{\Psi^L}$ and $\ket{\Psi^R}$
that are purely left- (contracting) and right- (expanding) moving in the
volume representation.  We first note that in the quantum theory the left- and
right- moving sectors are superselected by the Dirac observables, \emph{i.e.\
}the action of Dirac observables does not mix these sectors (see
Ref.\ \cite{aps} for details.)  From Eq.\ (\ref{eq:quant_traj}), we know that
$\Psi^L(\nu_+)=\bracket{\nu}{\Psi^L(\phi)}$ and
$\Psi^R(\nu_-)=\bracket{\nu}{\Psi^R(\phi)}$ only depend on $\nu$ and $\phi$ in
the combination $\nu_{\pm}=\phi\pm (12 \pi G)^{-1/2}\ln\nu$.  For any fixed
$\phi$, the range $\Delta\nu^*=[0,\nu^*]$ corresponds to
$\Delta\nu_{\pm}^*=(\mp\infty,\nu_{\pm}^*]$, and with $d\nu_{\pm}=\pm (12 \pi
G)^{-1/2}d\nu/\nu$,
\begin{subequations}
\begin{eqnarray}
p^L_{\Delta\nu^*}(\phi)
& = & \int_{0}^{\nu^*}\!\!\frac{d\nu}{\nu}\,|\Psi^L(\nu_+)|^2
\label{eq:pvolL-a}\\
& = & \k\, \int_{-\infty}^{\phi+\k^{-1}\ln\nu^*}\!\!d\nu_+\ |\Psi^L(\nu_+)|^2,
\label{eq:pvolL-b}
\end{eqnarray}
\label{eq:pvolL}%
\end{subequations}
and similarly
\begin{subequations}
\begin{eqnarray}
p^R_{\Delta\nu^*}(\phi)
& = & -\k\,  \int_{\infty}^{\phi-\k^{-1}\ln\nu^*}\!\!d\nu_-\ |\Psi^R(\nu_-)|^2,
\label{eq:pvolR-a}\\
& = & \k\, \int_{\phi-\k^{-1}\ln\nu^*}^{\infty}\!\!d\nu_-\ |\Psi^R(\nu_-)|^2
\label{eq:pvolR-b}
\end{eqnarray}
\label{eq:pvolR}%
\end{subequations}
where we have defined $\k = \sqrt{12 \pi G}$.

From Eqs.\ (\ref{eq:pvolL})-(\ref{eq:pvolR}) it is clear that since
$\ket{\Psi^{L,R}}$ are normalized states, 
\begin{subequations}
\begin{align}
\lim_{\phi\rightarrow-\infty} p^L_{\Delta\nu^*}(\phi) &= 0
&
\lim_{\phi\rightarrow+\infty} p^L_{\Delta\nu^*}(\phi) &= 1
\label{eq:label-a}\\
\lim_{\phi\rightarrow-\infty} p^R_{\Delta\nu^*}(\phi) &= 1
&
\lim_{\phi\rightarrow+\infty} p^R_{\Delta\nu^*}(\phi) &= 0
\label{eq:label-b}
\end{align}
\label{eq:label}%
\end{subequations}
for any fixed choice of $\nu^*$, no matter how small.

These are precisely the results we should have expected: contracting universes
will inevitably shrink to arbitrarily small volume, and expanding universes
have inevitably grown \emph{from} arbitrarily small volume.  In this sense,
then, purely expanding or contracting universes are inevitably singular at
either $\phi=-\infty$ or $+\infty$.

While this may have been the expected result, a few points are worth emphasis.
First, this result has been \emph{rigorously derived} according to an explicit
criterion within a fully coherent framework for deducing quantum
probabilities.  Second, while our intuition may be happy with this result for
classical or semiclassical universes, \emph{we have made no such assumption}
concerning the states $\ket{\Psi^{L}}$ and $\ket{\Psi^{R}}$.  Indeed, we made
no assumption at all about these states other than that they are purely left-
or right-moving.%
\footnote{Note that it was established in Ref.\ \cite{acs:slqc} that the
\emph{expectation value} of the volume hits zero (i) for generic left-moving
states as $\phi \rightarrow \infty$, and (ii) for generic right-moving states
as $\phi \rightarrow -\infty$ .  However, a consistent quantum probabilistic
interpretation was lacking.  Moreover, the present framework makes it possible
to ask and answer much more precisely framed questions.  Here, for example, we
find the probability for the quantum universe to be found in a specific range
of volumes -- not simply calculate its average value.  To illustrate why this
is significant, note that the expectation value of the volume in a state which
is a superposition of left- and right-moving modes (as in Eq.\
(\ref{eq:volcat}) below) will not asymptote to zero in either direction
$\phi\rightarrow\pm\infty$.  The expectation value cannot reveal that such
states are nonetheless invariably singular -- it is too coarse a diagnostic.
For that, one requires a consistent quantum analysis of ranges of possible
values of the volume at (at least) \emph{two} values of $\phi$, as discussed
below.
} %

What about more general states which are \emph{superpositions} of left- and
right-moving states?  Indeed, a glance at Fig.\ \ref{fig:classicalcg} might
lead one to wonder whether such a superposition might lead to the possibility
of a ``quantum bounce'', in which a superposition of (possibly macroscopic)
expanding and contracting universes might be likely to be peaked on a large
classical solution at \emph{both} ``early'' and ``late'' values of $\phi$ --
effectively ``jumping'' from one branch to the other.
We shall show that in these models, this cannot occur.  

\begin{figure}[hbt!]
\includegraphics[scale=0.84]{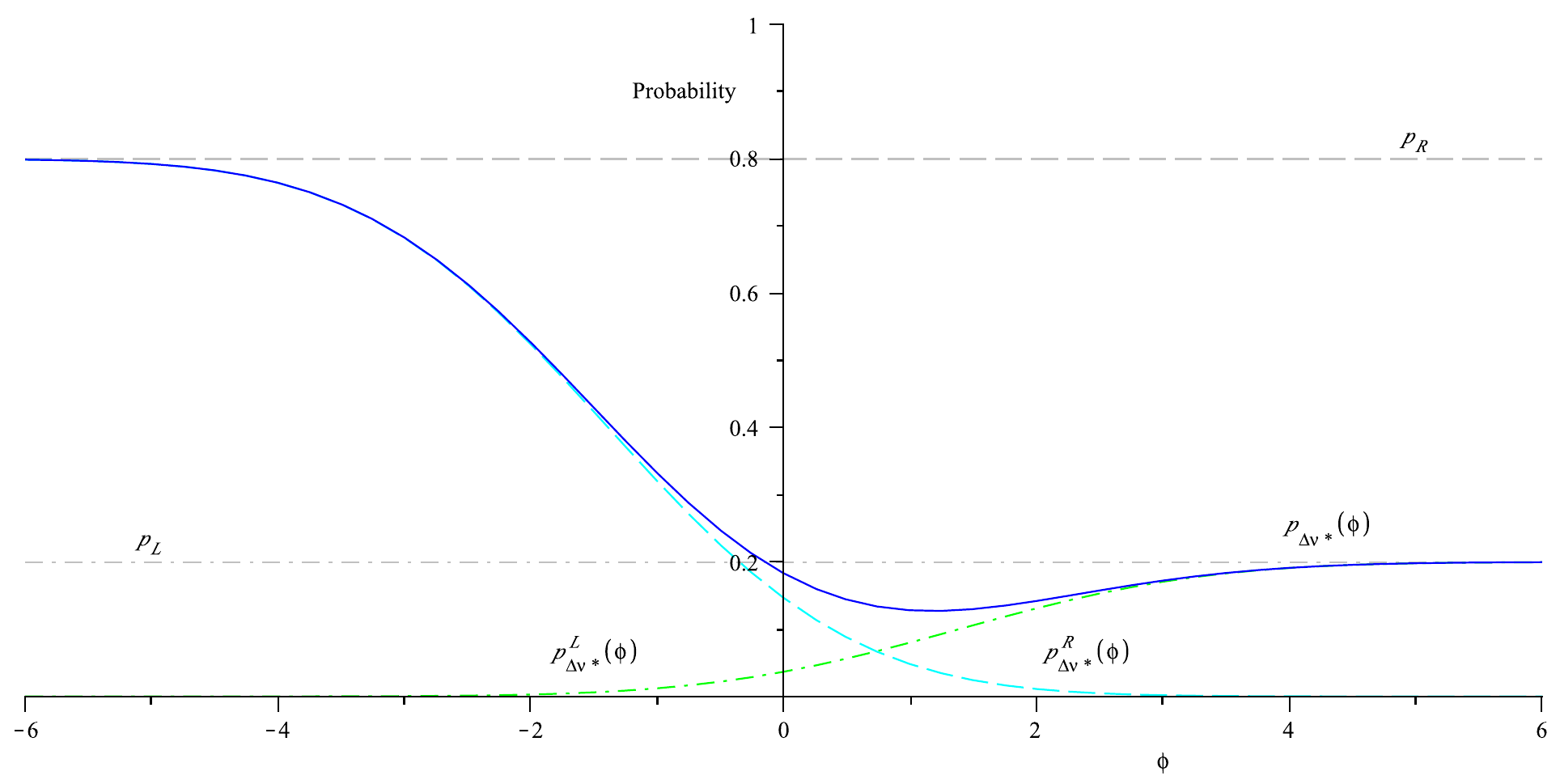}
\caption{The behavior of $p_{\Delta\nu^*}(\phi)$, the probability that the
quantum universe will be found in the interval $\Delta\nu^* =[0,\nu^*]$
(\emph{i.e.\ }at small volume) for a sample superposition of expanding $(R)$
and contracting $(L)$ semiclassical states both peaked at large volume near
$\phi=0$ (\emph{cf.\ }Eqs.\ (\ref{eq:semiclass}) and (\ref{eq:pvolsc}).)
$p_L$ and $p_R$ give the relative ``amount'' of each component in the
superposition, so that $p_L + p_R =1$.  (For the states shown in the figure,
$\phi_o=0$, $\bar{\nu}=10,000$, $\sigma=0.075$, $p_L = 0.2$, $\nu^*=2$, and we
have set $\lp =1$.)  Also shown are $p_{\Delta\nu^*}^L(\phi)$ and
$p_{\Delta\nu^*}^R(\phi)$, the probabilities that each component individually
will be found at small volume.  Note how $p_{\Delta\nu^*}^L(\phi)$ asymptotes
to $p_L$ as $\phi\rightarrow +\infty$ and $p_{\Delta\nu^*}^R(\phi)$ asymptotes
to $p_R$ as $\phi\rightarrow -\infty$, since the contracting component becomes
singular for large $\phi$, and the expanding component becomes singular for
large negative $\phi$.  Similarly, each component asymptotes to 0 in the
opposite direction.  This plot may appear to imply the possibility of a
``quantum bounce'', since at any given $\phi$ there is a non-zero probability
that the universe may be found with volume $\nu > \nu^*$.  A more careful
consistent histories analysis shows that this possibility is not realized: the
probability that the universe has large volume in \emph{both} the ``past'' and
``future'' is zero.  (After Ref.\ \cite{CS10b}.)
} %
\label{fig:pvolplot}
\end{figure}

To see how this works, consider the superposition
\begin{equation}
\ket{\Psi} = \ket{\Psi^L} + \ket{\Psi^R}.
\label{eq:volcat}
\end{equation}
Note, though, the states $\ket{\Psi^{L,R}}$ are no longer normalized.   
Rather
\begin{subequations}
\begin{eqnarray}
\bracket{\Psi}{\Psi} & = & \bracket{\Psi^L}{\Psi^L} + \bracket{\Psi^R}{\Psi^R}
\label{eq:volcatnorm-a}\\
& \equiv & p_L + p_R
\label{eq:volcatnorm-b}\\
& = & 1
\label{eq:volcatnorm-c}
\end{eqnarray}
\label{eq:volcatnorm}%
\end{subequations}
since $\ket{\Psi}$ is normalized and the left- and right-moving sectors are
orthogonal.  $p_L$ and $p_R$ measure the ``amount'' of each state in
$\ket{\Psi}$.  We now find that, since the volume projections leave the $L$
and $R$ sectors invariant,
\begin{equation}
p_{\Delta\nu^*}(\phi) = 
  p^L_{\Delta\nu^*}(\phi) + p^R_{\Delta\nu^*}(\phi),
\label{eq:pvolcat}
\end{equation}
with $p^{L,R}_{\Delta\nu^*}(\phi)$ given again by Eqs.\ 
(\ref{eq:pvolL})-(\ref{eq:pvolR}).  
Then
\begin{equation}
\lim_{\phi\rightarrow -\infty} p_{\Delta\nu^*}(\phi) = p_R
\qquad \mathrm{and}\qquad
\lim_{\phi\rightarrow +\infty} p_{\Delta\nu^*}(\phi) = p_L.
\label{eq:pvolcatlim}
\end{equation}
In between, we may expect $p_{\Delta\nu^*}(\phi)$ to drop even to very small
values, especially for $\ket{\Psi^{L,R}}$ that are peaked on classical
trajectories.  By way of example, for a left- and right-moving superposition
of the semiclassical states given in Eq.\ (\ref{eq:semiclass}), from Eqs.\ 
(\ref{eq:pvolsc-sm}) and (\ref{eq:pvolcat}) we find
%
%
\begin{equation}
p_{\Delta\nu^*}(\phi) = 
\frac{1}{2}\left\{1 + 
p_L\erf(\sigma_L(\ln\frac{\nu^*}{\bar{\nu}_L} + \cons(\phi-\phi_o))) +
p_R\erf(\sigma_R(\ln\frac{\nu^*}{\bar{\nu}_R} - \cons(\phi-\phi_o)))
\right\}.
\label{eq:pvolcatsc}
\end{equation}
%
The behaviour of $p^L_{\Delta\nu^*}(\phi)$, $p^R_{\Delta\nu^*}(\phi)$, and
$p_{\Delta\nu^*}(\phi)$ is shown in Fig.\ \ref{fig:pvolplot}.

\subsubsection{Absence of Quantum Bounce}
\label{sec:bounce}

It may appear that this result leaves open the possibility of a quantum bounce
for a universe in such a superposition, for at no value of $\phi$ does the
probability that the volume of the universe goes to zero become in general close
to unity.  At any given value of $\phi$, there is some probability for the
universe to assume small volume, and some probability for it to be large.  Is
there not then a non-zero probability for the universe to be large in
\emph{both} the ``past'' and the ``future''?  This is not the case, however,
in part because $p_{\Delta\nu^*}(\phi)$ is the \emph{wrong question to ask.}

This is so because it only inquires about the volume \emph{at a single value
of $\phi$.} Indeed, for a universe which is a superposition of both expanding
and contracting components, it is not surprising that at any given value of
$\phi$ -- including infinitely far in the ``past'' or ``future'' -- there is a
possibility the universe may be found in the ``other'' state, the one that is
not at arbitrarily small volume at that value of $\phi$.  For a ``quantum
bounce'' there must be a possibility that the universe is large at \emph{two}
values of $\phi$, both ``early'' and ``late''.  In other words, to properly
address the question of a quantum bounce, we must consider histories of the
form Eq.\ (\ref{eq:classopvolseq}) for two values of $\phi$.  This is where
the role of decoherence will be essential.

In particular, we must ask the question, ``What is the probability
that the universe was \emph{never} singular (had zero volume)?'', and
the complementary question, ``What is the probability that the
universe was \emph{ever} singular (had zero volume)?''  We shall find
that this family of histories is consistent, and that the answer to
the former question is zero, and the latter, unity.

The class operators for this coarse-graining are more subtle than one may at
first expect.  For concreteness let us consider a choice of two $\phi$-slices,
$\phi_1$ and $\phi_2$, and the corresponding partitions
$(\Delta\nu^*_1,\overline{\Delta\nu^*_1})$ and
$(\Delta\nu^*_2,\overline{\Delta\nu^*_2})$.  The choice of these slices is
such that we can probe the physics for both late and early times by taking for
example $\phi_1 \rightarrow -\infty$ and $\phi_2 \rightarrow \infty$.  The
class operator for the (very coarse-grained version of the) history ``the
universe has large volume at early as well as late times'' (\emph{i.e.\ }is
found in both $\overline{\Delta\nu^*_1}$ and $\overline{\Delta\nu^*_2}$) is
\begin{subequations}
\begin{eqnarray}
C_{\mathrm{bounce}}(\phi_1,\phi_2) &=&
C_{\overline{\Delta\nu^*_1};\overline{\Delta\nu^*_2}}
\label{eq:Cnever-a}\\
& = & 
\Projsupb{\nu}{\overline{\Delta\nu^*_1}}(\phi_1)
\Projsupb{\nu}{\overline{\Delta\nu^*_2}}(\phi_2).
\label{eq:Cnever-b}
\end{eqnarray}
\label{eq:Cnever}%
\end{subequations}
Since the class operators for an exclusive, exhaustive set of histories must 
sum to unity,
\begin{equation}
1 = 
C_{\smash{\Delta\nu^*_1;\Delta\nu^*_2}} + 
C_{\smash{\Delta\nu^*_1;\overline{\Delta\nu^*_2}}} +
C_{\smash{\overline{\Delta\nu^*_1};\Delta\nu^*_2}} +
C_{\smash{\overline{\Delta\nu^*_1};\overline{\Delta\nu^*_2}}},
\label{eq:Csum}
\end{equation}
the class operator for the complementary history ``the universe is not
found in both
$\overline{\Delta\nu^*_1}$ and
$\overline{\Delta\nu^*_2}$'', or in other words, ``the universe will be in 
$\Delta\nu^*_1$ (and) or
$\Delta\nu^*_2$,'' is given by
\begin{subequations}
\begin{eqnarray}
C_{\mathrm{sing}}(\phi_1,\phi_2)  & = & 
1 - C_{\overline{\Delta\nu^*_1};\overline{\Delta\nu^*_2}}
\label{eq:Cever-a}\\
& = & 
C_{\smash{\Delta\nu^*_1;\Delta\nu^*_2}} + 
C_{\smash{\Delta\nu^*_1;\overline{\Delta\nu^*_2}}} +
C_{\smash{\overline{\Delta\nu^*_1};\Delta\nu^*_2} }
\label{eq:Cever-b}\\
& = & 
C_{\smash{\Delta\nu^*_1;\Delta\nu^*_2\cup\overline{\Delta\nu^*_2}}} + 
C_{\smash{\Delta\nu^*_1\cup\overline{\Delta\nu^*_1};\Delta\nu^*_2}} -
C_{\smash{\Delta\nu^*_1;\Delta\nu^*_2}} 
\label{eq:Cever-c}\\
& = & 
C_{\Delta\nu^*_1} + 
C_{\Delta\nu^*_2} -
C_{\Delta\nu^*_1;\Delta\nu^*_2},
\label{eq:Cever-d}
\end{eqnarray}
\label{eq:Cever}%
\end{subequations}
where it should be clear that, for example,
$C_{\smash{\Delta\nu^*_1;\Delta\nu^*_2}} +
C_{\smash{\Delta\nu^*_1;\overline{\Delta\nu^*_2}}} =
C_{\smash{\Delta\nu^*_1;\Delta\nu^*_2\cup\overline{\Delta\nu^*_2}}} =
C_{\Delta\nu^*_1}$, since the $C_{\Delta\nu_1;\Delta\nu_2}$ are simply
products of projections.  One can see that the terms on the right hand side of
Eq.\ (\ref{eq:Cever-b}) encode the various ways in which the universe may find
itself at small volume at one or both of $\phi_1$ and $\phi_2$.

We now show that $C_{\mathrm{bounce}}^{\dagger}\ket{\Psi}=0$ if we take
$\phi_1\rightarrow -\infty$ and $\phi_2\rightarrow +\infty$, and therefore the
set of histories (bounce, singular) decoheres, with $p_{\mathrm{sing}}=1$.

We will proceed by demonstrating that 
$C_{\smash{\Delta\nu^*_1;\Delta\nu^*_2}}^{\dagger}\ket{\Psi}$ and
$C_{\smash{\overline{\Delta\nu^*_1};\overline{\Delta\nu^*_2}}}^{\dagger}\ket{\Psi}$
both vanish in this limit.   To see how this works, let us calculate
$\Projsupb{\nu}{\Delta\nu}(\phi)\ket{\Psi^L}$, again employing 
$\nu_{\pm}=\phi\pm\k^{-1}\ln\nu$:
\begin{subequations}
\begin{eqnarray}
\Projsupb{\nu}{\Delta\nu}(\phi)\ket{\Psi^L}
& = & 
U^{\dagger}(\phi-\phi_0)\Projsupb{\nu}{\Delta\nu}\ket{\Psi^L(\phi)}
\label{eq:projvolpsi-a}\\
& = & 
U^{\dagger}(\phi-\phi_0)\int_{\Delta\nu}\!\!\frac{d\nu}{\nu}\ket{\nu}\Psi^L(\nu,\phi)
\label{eq:projvolpsi-b}\\
& = & 
U^{\dagger}(\phi-\phi_0)\k\, 
  \int_{\Delta\nu_+}\!\!\!d\nu_+\ket{\nu_+}\Psi^L(\nu_+),
\label{eq:projvolpsi-c}
\end{eqnarray}
\label{eq:projvolpsi}%
\end{subequations}
where we have introduced 
$\ket{\nu_{\pm}}\equiv\ket{\nu=\exp(\pm\k(\nu_{\pm}-\phi))}$, %
which satisfy
$\bracket{\nu_{\pm}'}{\nu_{\pm}}=\k^{-1}\delta(\nu_{\pm}'-\nu_{\pm})$ %
and
$1=\k\int_{-\infty}^{\infty}d\nu_{\pm}\ketbra{\nu_{\pm}}{\nu_{\pm}}$. %

Thus we see that
\begin{subequations}
\begin{eqnarray}
\lim_{\phi\rightarrow\pm\infty}\Projsupb{\nu}{\Delta\nu^*}(\phi)\ket{\Psi^L}
& = & 
\lim_{\phi\rightarrow\pm\infty}  U^{\dagger}(\phi-\phi_0)\, \k\, 
  \int_{-\infty}^{\phi+\k^{-1}\ln\nu^*}\!\!\!d\nu_+\ket{\nu_+}\Psi^L(\nu_+),
\label{eq:projvolsingL-a}\\
& = & 
\lim_{\phi\rightarrow\pm\infty}  U(\phi_0-\phi)\,
\left\{
\begin{array}{l}   
         \ket{\Psi^L(\phi)}     \\
         0                                             
\end{array}\right.  
\label{eq:projvolsingL-b}\\
& = & 
\left\{
\begin{array}{lcl}   
         \ket{\Psi^L}  && \phi\rightarrow +\infty   \\
         0             && \phi\rightarrow -\infty                          
\end{array}\right.  
.
\label{eq:projvolsingL-c}
\end{eqnarray}
\label{eq:projvolsingL}%
\end{subequations}
Similarly,
\begin{equation}
\lim_{\phi\rightarrow\pm\infty}\Projsupb{\nu}{\Delta\nu^*}(\phi)\ket{\Psi^R}
=
\left\{
\begin{array}{lcl}   
         0             && \phi\rightarrow +\infty   \\
         \ket{\Psi^R}  && \phi\rightarrow -\infty                          
\end{array}\right.  
\label{eq:projvolsingR}
\end{equation}
and in the same way
\begin{equation}
\lim_{\phi\rightarrow\pm\infty}\Projsupb{\nu}{\overline{\Delta\nu^*}}(\phi)\ket{\Psi^L}
=
\left\{
\begin{array}{lcl}   
         0             && \phi\rightarrow +\infty   \\
         \ket{\Psi^L}  && \phi\rightarrow -\infty                          
\end{array}\right.  
\label{eq:projvolbounceL}
\end{equation}
and
\begin{equation}
\lim_{\phi\rightarrow\pm\infty}\Projsupb{\nu}{\overline{\Delta\nu^*}}(\phi)\ket{\Psi^R}
=
\left\{
\begin{array}{lcl}   
         \ket{\Psi^R}  && \phi\rightarrow +\infty   \\
         0             && \phi\rightarrow -\infty                          
\end{array}\right.  
.
\label{eq:projvolbounceR}
\end{equation}

In this way, given the general superposition Eq.\ (\ref{eq:volcat}),
\begin{subequations}
\begin{eqnarray}
\lim_{\substack{\phi_1\rightarrow -\infty\\ \phi_2\rightarrow +\infty }}
C_{\smash{\Delta\nu^*_1;\Delta\nu^*_2}}^{\dagger}\ket{\Psi}
& = & 
\lim_{\substack{\phi_1\rightarrow -\infty\\ \phi_2\rightarrow +\infty }}
\Projsupb{\nu}{\Delta\nu^*}(\phi_2)
\Projsupb{\nu}{\Delta\nu^*}(\phi_1)
\left\{\ket{\Psi^L}+\ket{\Psi^R}\right\}
\label{eq:projvolsingsing-a}\\
& = & 
\lim_{\phi_2\rightarrow +\infty}
\Projsupb{\nu}{\Delta\nu^*}(\phi_2)
\ket{\Psi^R}
\label{eq:projvolsingsing-b}\\
& = & 
0,
\label{eq:projvolsingsing-c}
\end{eqnarray}
\label{eq:projvolsingsing}%
\end{subequations}
and in the same way
\begin{equation}
\lim_{\substack{\phi_1\rightarrow -\infty\\ \phi_2\rightarrow +\infty }}
C_{\smash{\overline{\Delta\nu^*_1};\overline{\Delta\nu^*_2}}}^{\dagger}\ket{\Psi} =0.
\label{eq:projvolbouncebounce}
\end{equation}
Defining formally
\begin{equation}
C_{\mathrm{bounce}} = 
\lim_{\substack{\phi_1\rightarrow -\infty\\ \phi_2\rightarrow +\infty }}
C_{\mathrm{bounce}}(\phi_1,\phi_2)
\label{eq:Cbounce}
\end{equation}
and
\begin{equation}
C_{\mathrm{sing}} = 
\lim_{\substack{\phi_1\rightarrow -\infty\\ \phi_2\rightarrow +\infty }}
C_{\mathrm{sing}}(\phi_1,\phi_2),
\label{eq:Csing}
\end{equation}
where $C_{\mathrm{bounce}}(\phi_1,\phi_2)$ and
$C_{\mathrm{sing}}(\phi_1,\phi_2)$ are given in Eqs.\
(\ref{eq:Cnever}) and (\ref{eq:Cever}), we see that the branch wave functions are
\begin{subequations}
\begin{eqnarray}
\ket{\Psi_{\mathrm{bounce}}} & = & C_{\mathrm{bounce}}^{\dagger}\ket{\Psi}
\label{eq:psibounce-a}\\
& = & 0
\label{eq:psibounce-b}
\end{eqnarray}
\label{eq:psibounce}%
\end{subequations}
and
\begin{subequations}
\begin{eqnarray}
\ket{\Psi_{\mathrm{sing}}} & = & C_{\mathrm{sing}}^{\dagger}\ket{\Psi}
\label{eq:psising-a}\\
& = & \ket{\Psi^L}+\ket{\Psi^R}
\label{eq:psising-b}\\
& = & \ket{\Psi}.
\label{eq:psising-c}
\end{eqnarray}
\label{eq:psising}%
\end{subequations}
Since only one of the branch wave functions is not zero, the histories 
(bounce,singular) clearly decohere, and
\begin{subequations}
\begin{eqnarray}
p_{\mathrm{sing}} & = & \bracket{\Psi_{\mathrm{sing}}}{\Psi_{\mathrm{sing}}}
\label{eq:psing-a}\\
& = & \bracket{\Psi}{\Psi}
\label{eq:psing-b}\\
& = & 1.
\label{eq:psing-c}
\end{eqnarray}
\label{eq:psing}%
\end{subequations}
Note once again that we have made no assumptions on the state whatever
-- this result holds for all choices of state, highly quantum, highly
classical, or even a Schr\"{o}dinger cat-like superposition of
expanding and collapsing universes.  All quantum Wheeler-DeWitt
universes will assume zero volume at some point in their history.
There is no ``quantum bounce''.

It is important to emphasize that the question of whether a quantum universe
can ``bounce'' is \emph{not} trivial and can \textbf{not} be answered simply
by examining the behaviour of $\ket{\Psi(\phi)}$.  As emphasized in Ref.\
\cite{CS10a}, it is inherently a \emph{quantum} question, and the role of
decoherence is essential to its coherent analysis.  This is because it is
inherently a question about the value of the volume on two different
$\phi$-slices.  In general such histories
do \emph{not} decohere,%
\footnote{Even for purely left- or right-moving states -- never mind 
superpositions such as Eq.\ (\ref{eq:volcat}).
} %
and the question of what is the volume at two different values of $\phi$ makes
in general no more quantum sense than the question of which slit a particle
passed through in the case of two-slit
interference.%
\footnote{As also in that case, one might expect coupling to other degrees of
freedom to lead to decoherence
\cite{halliwell89,padman89,habiblaflamme90,bk90,*bk95}.
} %
Indeed, it is only in the limit that $\phi_1\rightarrow
-\infty$ and $\phi_2\rightarrow +\infty$ that we are assured in general that
these histories do decohere; that the quantum question may be answered, in the
sense that probabilities may be assigned; and that the answer turns out to be
that, in this quantization, these model universes never ``bounce''.

We wish to be clear what is being asserted here.  In tabletop quantum theory
the proper manner in which to address a question of this kind is to calculate
a \emph{conditional} probability (transition amplitude) -- \emph{if} the
universe is found to be at large volume in the ``past'', what is the
probability that it will \emph{also} be at large volume in the ``future''?
That is essentially what Eq.\ (\ref{eq:projvolbouncebounce}) determines.  The
utility of such probabilities on the tabletop, however, is predicated on the
assumption that the system has been \emph{measured} to be at large volume in
the past.  For a closed system (such as the universe), however, external
measurements do not exist, and the consistency of conditional probabilities
calculated in this way can only be assured by decoherence of the corresponding
histories.  In other words, \emph{physically meaningful probabilities cannot
be inferred from transition amplitudes unless the corresponding family of
histories is consistent.} In the case of the present example, for instance,
the probabilities so calculated are \textbf{only} consistent in the limiting
case of $\phi\rightarrow\pm\infty$.



\section{Discussion}
\label{sec:discuss}

An important conceptual issue in the application of quantum theory to the
whole universe is to understand the way in which quantum probabilities can be
assigned consistently to various phenomena.  Since by definition the universe
as a whole is a closed quantum system, one lacks the notion of external
observers/apparatus which can be treated classically to ``measure'' the
wavefunction of the universe.  The fundamental inadequacy of the Copenhagen
interpretation becomes clearly evident when applying quantum theory to the
cosmos.  The consistent histories framework in the form of Hartle's
generalized quantum mechanics has been advocated precisely to answer these
questions for a quantum universe.  In this work we have given an explicit
example of a Wheeler-DeWitt quantum universe where non-trivial questions about
decoherence between alternate histories can be posed and answered and
probabilities for various events be computed consistently and rigorously.  To
our knowledge this work is the first of its kind where consistent probabilities
for the occurrence of a singularity in a quantum gravitational model have been
calculated without any assumption on the states in the physical Hilbert space.

We have considered the Wheeler-DeWitt quantization of a flat homogeneous and
isotropic universe with a free, massless, minimally coupled scalar field.
Though the model is simple, it is non-trivial enough to pose interesting
questions.  Classically the model has two classes of solutions -- expanding
with a big bang singularity in the past, and contracting with big crunch
singularity in the future.  The physical Hilbert space, Dirac observables and
their expectation values for this model have been studied recently
\cite{aps,aps:improved,acs:slqc}.  The latter importantly showed that generic
left- and right-moving Wheeler-DeWitt universes are singular; we extend that
result to include superpositions of such states.  Using the basic apparatus
from these works we have completed its consistent histories analysis and
extracted quantum probabilities.  In contrast to interpreting the amplitude of
the wavefunction heuristically as in previous work on this model, we have
succeeded in rigorously \emph{deriving} explicit formul\ae\ for the quantum
probabilities within a consistent framework.  With these we were able to
answer questions such as ``What is the probability that a Wheeler-DeWitt
universe ever hits the singularity?''  Though one expected the answer to be
unity (from intuition of the behavior of the wavefunction in open quantum
systems,) it is here for the first time an explicit, internally consistent
computation of the probability, taking proper care of the consistency of the
corresponding quantum histories, has been shown to give this result.

An interesting example of such a calculation concerned a quantum state which
is a superposition of expanding and contracting universes.  Such a state might
be regarded as an analog of ``Schr\"{o}dinger's cat'' in quantum cosmology --
a superposition of (potentially) macroscopically distinct states.  For such
states one may be tempted to ask: Because there is an amplitude for the
universe to have large volume in the ``past'' ($\phi\rightarrow -\infty$), and
an amplitude for the universe to have large volume in the ``future''
($\phi\rightarrow +\infty$), is there not then an amplitude for the universe
to have large volume in \emph{both} the ``past'' and the ``future''?  Were it
so, one would have shown that for such states there is an amplitude for
Wheeler-DeWitt-quantized universes to avoid the cosmological singularity with
a ``quantum bounce'' similar to that which appears in models of loop quantum
cosmology \cite{aps,aps:improved,acs:slqc}.  As we have demonstrated, however,
a careful consistent histories analysis shows that the answer to this question
is \textbf{``No''}.  As emphasized in Ref.\ \cite{CS10a}, a ``quantum bounce''
is inherently a quantum question involving (at least) two slices of ``time'',
and only has a meaningful quantum answer if the corresponding histories
decohere.  We show, without making \emph{any} assumption on the nature of the
quantum state, that the corresponding branch wave functions \emph{do}
decohere, and the probability for such a bounce is zero.

A feature of this analysis was the notion of an emergent time derived from one
of the degrees of freedom in the phase space.  Here this was taken to be the
value of the scalar field, though the scale factor or volume of the universe
would have served just as well.  This enabled us to pose meaningful questions
about expectation values of relational observables -- \emph{viz.\ }the volume
of the universe at a given value of $\phi$, and the momentum of the scalar
field, and their probabilities.  We found that the probabilities computed from
class operators for quantum histories are consistent with expectation values
of the Dirac (relational) observables, thus pointing to an overall coherence
of the frameworks -- the consistent histories approach applied to canonical
quantization \emph{vs.\ }an analysis purely in terms of relational
observables.  (We hope to explore further details of this relationship in a
future work.)  The consistent histories framework, however, enables a
significantly more finely-grained study -- which, we emphasize, can ask and 
answer questions simple study of expectation values can\emph{not}.

While the model studied here is very simple, one may hope to employ the
framework for construction of the decoherence functional developed in Refs.\
\cite{hartle91a,lesH,CH04,halliwell09} and here as a template for the
consistent histories formulation of more sophisticated quantum gravitational
models.  Let us now point out the minimum features a model of quantum
cosmology in the canonical setting should have for the present analysis to be
extendible.  One of the key requirements is the availability of the physical
Hilbert space, inner product, and dynamics.  One must also have available at
least one non-trivial observable which can be consistently defined.
If the corresponding ``relational'' Dirac observable can be constructed which
correlates values of that observable with other degrees of freedom, it should
be possible to frame a coherent consistent histories formulation in the same
manner.  In particular, it is not essential that the degrees of freedom used
to define those correlations be monotonic, as they were in the present
example.  The presence of an ``emergent time'', while convenient, is
\emph{not} essential to the formulation of a consistent histories framework.
What matters is the ability to define correlations among degrees of freedom.

The analysis presented will be extended to study probabilities for the
(absence of) singularities in loop quantum cosmology in \cite{CS10d}.  For the
same model as considered here but in a loop quantization, we will show that
the answers turn out to be strikingly different.  We will demonstrate within
the consistent histories framework that a loop quantum universe never
encounters a singularity.  The probability that the universe bounces turns out
to be unity, thus providing a rigorous and consistent quantum probabilistic
interpretation to the results so far inferred from the behavior of expectation
values of Dirac observables.





\begin{acknowledgments}

D.C.\ would like to thank the Perimeter Institute, where much of this work was
completed, for its repeated hospitality.  D.C.\ would also like to extend his
gratitude to Helen Clark for her always gracious hospitality, which enabled
significant portions of the writing of this work to be undertaken.  P.S.\ is
grateful to Abhay Ashtekar for the suggestion to investigate the consistent
histories formalism.  Research at the Perimeter Institute is supported by the
Government of Canada through Industry Canada and by the Province of Ontario
through the Ministry of Research and Innovation.

\end{acknowledgments}


\bibliography{dWdW}

\end{document}